\documentclass[aps,pra,preprint,superscriptaddress,longbibliography]{revtex4-2}

\usepackage{amsmath,amsthm,amsfonts,amssymb}
\usepackage{bm}
\usepackage{microtype}
\usepackage{xcolor,graphicx}
\usepackage[pdftex,colorlinks=true,linkcolor=blue,citecolor=blue,urlcolor=blue]{hyperref} %para ligar las referencias

\begin{document}
	
	\title{Supersymmetry of Helmholtz waves}
	
	\author{{A.} {Dom\'inguez-Cruz}}% ORCID: 0009-0000-4485-8744
	\email[e-mail: ]{agustin.dominguez.cruz@gmail.com}
	\affiliation{Escuela de Ingenier\'ia y Ciencias, Tecnol\'{o}gico de Monterrey, Monterrey 64700, M\'{e}xico.}
	
	\author{{G.} {Badu}}
	\email[e-mail: ]{225194242@stu.ukzn.ac.za}
	\affiliation{School of Chemistry and Physics, University of KwaZulu-Natal, Private Bag X54001, Durban 4000, South Africa.}
	
	\author{{B.} {Perez-Garcia}}% ORCID: 0000-0002-5501-6291
	\email[e-mail: ]{perezgarciab@ukzn.ac.za}
	\affiliation{School of Chemistry and Physics, University of KwaZulu-Natal, Private Bag X54001, Durban 4000, South Africa.}
	
	\author{{B.~M.} {Rodr\'iguez-Lara}}% ORCID: 0000-0002-7014-0450
	\email[e-mail: ]{blas.rodriguez@gmail.com}
	\affiliation{Universidad Polit\'ecnica Metropolitana de Hidalgo, Tolcayuca, Hidalgo 43860, Mexico.}

	\begin{abstract}
		Symmetry organizes physical laws, links conservation principles to transformations, and reveals connections between seemingly different systems.
		We show that Helmholtz waves support Witten supersymmetry through their spatial symmetries, without additional internal degrees of freedom.
		Reflection across the transverse horizontal axis induces a $\mathbb{Z}_{2}$ grading of the Euclidean algebra, whose odd generators supply charges connecting opposite-parity Cartesian and Bessel modes at equal separation eigenvalues.
		The Bessel model has an unpaired even zero mode, Witten index one, and unbroken supersymmetry for any rotation-axis position along the reflection line.
		We express all four separators through mixed supercommutators of displaced orbital-angular-momentum generators, establish Weber pairing through a spectral sign operator, and identify the obstruction to equal-eigenvalue Mathieu pairing at nonzero ellipticity.
		We extend our construction to multiple transverse shells and generate supersymmetric partner fields with a common Talbot recurrence.
		Gaussian apodization centered on the rotation axis gives finite-power fields and preserves our charge relations during exact forward propagation while modifying the ideal revival.
		Our framework may guide optical transformations between parity partners and the characterization of structured fields.
	\end{abstract}

	\maketitle
	\clearpage
	
	%%%%%%%%%%%%%%%%%%%%%%%
	\section{Introduction}
	\label{sec:Sec1}
	%%%%%%%%%%%%%%%%%%%%%%%
	
	Control over the amplitude, phase, and polarization of light makes its spatial structure a resource for imaging, communication, and the manipulation of matter~\cite{Forbes2021p253}.
	An ideal electric field in a source-free homogeneous, isotropic, linear, lossless medium~\cite{Stratton1941},
	\begin{align}
		\bm{E} = \left( E_{\mathrm{TE}} \hat{\bm{M}} + E_{\mathrm{TM}} \hat{\bm{N}} \right) \Psi,
	\end{align}
	combines transverse-electric and transverse-magnetic contributions with modal coefficients $E_{\mathrm{TE}}$ and $E_{\mathrm{TM}}$~\cite{Rayleigh1897p125} and transverse-electric and transverse-magnetic field operators~\cite{RodriguezLara2008p033813,RodriguezLara2009055806},
	\begin{align}
		\begin{aligned}
			\hat{\bm{M}} \Psi =&~ - i k \bm{\nabla} \times \left( \bm{e}_{z} \Psi \right), \\
			\hat{\bm{N}} \Psi =&~ \bm{\nabla} \times \bm{\nabla} \times \left( \bm{e}_{z} \Psi \right),
		\end{aligned}
	\end{align}
	where $k = \omega / v$ is the medium wavenumber for angular frequency $\omega$ and phase velocity $v$, and $\bm{e}_{z}$ is the axial unit vector.
	For a fixed axial wavenumber $k_{z}$, the scalar wave function,
	\begin{align}
		\Psi\left( x,y,z,t \right) = \Phi\left( x,y \right) e^{i\left( k_{z} z - \omega t \right)},
	\end{align}
	has a transverse profile $\Phi \left( x, y \right)$ that satisfies the constant-shell transverse Helmholtz equation~\cite{Durnin1987p651},
	\begin{align}
		\left( \partial_{x}^{2} + \partial_{y}^{2} + k_{\perp}^{2} \right) \Phi = 0,
	\end{align}
	with fixed transverse wavenumber $k_{\perp} = \sqrt{k^{2} - k_{z}^{2}} > 0$.
	Its symmetries determine separation of variables in four orthogonal coordinate systems, Cartesian, circular, elliptic, and parabolic~\cite{Miller1974p626,Boyer1976p35}.
	Their separation operators combine transverse translations and axial rotations, commute with the constant-shell transverse Helmholtz operator, and have the separation constants as eigenvalues~\cite{Miller1977}.
	
	Cartesian plane waves provide the common Whittaker representation of these families~\cite{Whittaker1903p333}.
	Bessel experiments established approximately invariant profiles~\cite{Durnin1987p1499}, particle manipulation in multiple planes~\cite{GarcesChavez2002p145}, microscopy through scattering media~\cite{Fahrbach2010p780}, and thin-sheet imaging of living cells~\cite{Planchon2011p417}.
	High-order Bessel waves transfer orbital angular momentum (OAM) to trapped particles~\cite{VolkeSepulveda2002pS82}, and their quantization identifies the corresponding field observables~\cite{Jauregui2005p033411}.
	Frequency-conversion theory predicts radially polarized Bessel output~\cite{Saltiel2007p4132}, while digital axicons resolve Bessel mode content~\cite{Trichili2014p17553}.
	Mathieu waves extend propagation-invariant profiles to elliptic coordinates~\cite{GutierrezVega2000p1493} and support elliptic vortices~\cite{ChavezCerda2001p1803}.
	Plane-wave expansions determine their angular functions~\cite{ChaosCador2002p67}, and Bessel decompositions quantify the orbital angular momentum of experimentally generated Mathieu modes~\cite{ChavezCerda2002pS52}.
	Weber waves complete the separable families~\cite{Bandres2004p44}, with experiments resolving even, odd, and traveling fields and their transverse energy flow~\cite{LopezMariscal2005p2364}.
	Their mechanical effects include scattering of atomic clouds~\cite{HernandezCedillo2013p023404}, with parabolic momentum describing their conserved quantity and curved momentum flow~\cite{Bandres2013p013054}.
	Their normalization and Bessel and Mathieu decompositions~\cite{RodriguezLara2010p327} now extend to bidirectional Bessel--Weber expansions~\cite{DominguezCruz2026p055601} and Mathieu--Weber transformations~\cite{DominguezCruz2026ae929a}.
	
	In planar waveguides, the transverse profiles of guided modes satisfy a Helmholtz equation with an effective potential set by the refractive-index profile~\cite{Snyder1983}.
	Supersymmetry uses factorization to construct partner refractive-index profiles whose guided modes share propagation constants with the original system.
	For unbroken supersymmetry, one system has an unpaired ground mode, while for broken supersymmetry every mode has a partner.
	An early optical construction connected beams of different frequencies in the paraxial approximation~\cite{Chumakov1994p51}, while later work produced partner waveguides with shared propagation constants~\cite{Miri2013p233902} and demonstrated optical mode conversion~\cite{Heinrich2014p3698}.
	In waveguide arrays, factorization yields Jacobi partners with unbroken supersymmetry~\cite{ZunigaSegundo2014p987} and isospectral Cholesky partners with broken supersymmetry~\cite{MartinezBerumen2020p043521}.
	Applications include structures with identical reflection and transmission~\cite{Miri2014p89}, single-mode laser arrays~\cite{Hokmabadi2019p623}, and designed transmission resonances in photonic bandgaps~\cite{Chandra2021p59}.
	These constructions do not determine the pairing conditions imposed by spatial symmetry on the four separable families in a homogeneous medium.
	
	We construct a common supersymmetric framework for Helmholtz waves from their spatial symmetries and use its charges to generate partner fields.
	We review the literature on the Euclidean algebra of Helmholtz waves in Sec.~\ref{sec:Sec2} and their angular spectra in Sec.~\ref{sec:Sec3}.
	We use spatial reflection to grade the Euclidean generators and construct our superalgebra in Sec.~\ref{sec:Sec4}.
	Within this algebra, we introduce our displaced OAM generators and express the four separation operators through their mixed supercommutators in Sec.~\ref{sec:Sec5}.
	Using our grading, we construct supersymmetric charges and determine which modes they connect at the same separation eigenvalue in Sec.~\ref{sec:Sec6}.
	Extending these charges to multiple transverse shells in Sec.~\ref{sec:Sec7}, we generate partner sequences with a common paraxial Talbot recurrence and then compare their numerical and experimental propagation under centered Gaussian apodization in Sec.~\ref{sec:Sec8}.
	We close with our conclusions in Sec.~\ref{sec:Sec9}.
	
	%%%%%%%%%%%%%%%%%%%%%%%
	\section{Euclidean algebra}
	\label{sec:Sec2}
	%%%%%%%%%%%%%%%%%%%%%%%
	
	We summarize the Euclidean algebra and the four separable Helmholtz families using the dimensionless transverse coordinates~\cite{DominguezCruz2026p2606.26068},
	\begin{align}
		Q_{c} = k_{\perp} c,
	\end{align}
	with Cartesian coordinate label $c \in \{ x,y \}$ and position operators $\hat{Q}_{c}$ acting by multiplication by $Q_{c}$, leading to the transverse translation generators and axial orbital-angular-momentum generator~\cite{Miller1974p626},
	\begin{align}
		\begin{aligned}
			\hat{P}_{c} =&~ - i \partial_{Q_{c}}, \\
			\hat{L}_{z} =&~ \hat{Q}_{x} \hat{P}_{y} - \hat{Q}_{y} \hat{P}_{x},
		\end{aligned}
	\end{align}
	that close the Euclidean algebra $\mathrm{e}(2)$,
	\begin{align}
		\begin{aligned}
			\left[ \hat{L}_{z}, \hat{P}_{x} \right] =&~ i \hat{P}_{y}, \\
			\left[ \hat{L}_{z}, \hat{P}_{y} \right] =&~ - i \hat{P}_{x}, \\
			\left[ \hat{P}_{x}, \hat{P}_{y} \right] =&~ 0,
		\end{aligned}
	\end{align}
	with quadratic Casimir~\cite{Miller1977},
	\begin{align}
		\hat{P}^{2} = \hat{P}_{x}^{2} + \hat{P}_{y}^{2},
	\end{align}
	that commutes with all three generators,
	\begin{align}
		\left[ \hat{P}^{2}, \hat{P}_{x} \right] = \left[ \hat{P}^{2}, \hat{P}_{y} \right] = \left[ \hat{P}^{2}, \hat{L}_{z} \right] = 0,
	\end{align}
	and gives the unit-shell form of the transverse Helmholtz equation,
	\begin{align}
		\hat{P}^{2} \Phi = \Phi.
	\end{align}
	
	The Cartesian, Bessel, Mathieu, and Weber separation operators~\cite{Boyer1976p35},
	\begin{align}
		\begin{aligned}
			\hat{A}_{\mathrm{C}} =&~ \frac{1}{4} \left( \hat{P}_{x}^{2} - \hat{P}_{y}^{2} \right), \\
			\hat{A}_{\mathrm{B}} =&~ \hat{L}_{z}^{2}, \\
			\hat{A}_{\mathrm{M}} =&~ \hat{L}_{z}^{2} + \frac{1}{2} Q_{f}^{2} \left( \hat{P}_{x}^{2} - \hat{P}_{y}^{2} \right), \\
			\hat{A}_{\mathrm{W}} =&~ \frac{1}{2} \left( \hat{L}_{z} \hat{P}_{y} + \hat{P}_{y} \hat{L}_{z} \right),
		\end{aligned}
	\end{align}
	belong to the quadratic sector of the universal enveloping algebra of $\mathrm{e}(2)$ and commute with the Casimir,
	\begin{align}
		\left[ \hat{P}^{2}, \hat{A}_{\nu} \right] = 0,
	\end{align}
	where $f$ is the elliptic half-interfocal distance, $Q_{f} = k_{\perp} f$ is its dimensionless form, and $\nu \in \{ \mathrm{C},\mathrm{B},\mathrm{M},\mathrm{W} \}$ labels the family.
	Each separator shares generalized eigenfunctions with the Casimir, satisfying the separation eigenvalue equation on the unit Helmholtz shell,
	\begin{align}
		\hat{A}_{\nu} \Phi_{\lambda_{\nu}}^{\left( \nu \right)} = a_{\lambda_{\nu}}^{\left( \nu \right)} \Phi_{\lambda_{\nu}}^{\left( \nu \right)},
	\end{align}
	where the real transverse profiles with well defined parity~\cite{DominguezCruz2026p2606.26068,Durnin1987p651,RodriguezLara2008p033813,RodriguezLara2009055806},
	\begin{align}
		\begin{aligned}
			\Phi_{\bm{\sigma},\vartheta}^{\left( \mathrm{C} \right)}\left( Q_{x},Q_{y} \right) =&~ \sqrt{\frac{2}{\pi}} \, \mathcal{T}_{\sigma_{x}}\left( Q_{x} \cos \vartheta \right) \, \mathcal{T}_{\sigma_{y}}\left( Q_{y} \sin \vartheta \right), \\
			\Phi_{\sigma,\ell}^{\left( \mathrm{B} \right)}\left( Q_{\rho},Q_{\theta} \right) =&~ \sqrt{\frac{2}{1 + \delta_{\ell,0}}} \, J_{\ell}\left( Q_{\rho} \right) \, \mathcal{T}_{\sigma}\left( \ell Q_{\theta} \right), \\
			\Phi_{\sigma,m}^{\left( \mathrm{M} \right)}\left( Q_{\eta},Q_{\xi} \right) =&~ \sqrt{2} \, Y_{\sigma,m}\left( Q_{\eta} \right) \, M_{\sigma,m}\left( Q_{\xi} \right), \\
			\Phi_{\sigma,b}^{\left( \mathrm{W} \right)}\left( Q_{u},Q_{v} \right) =&~ \frac{\left\lvert \Gamma\left( \frac{1 + 2 \delta_{\sigma,\mathrm{o}} + 2 i b}{4} \right) \right\rvert^{2}}{2^{\delta_{\sigma,\mathrm{e}}} \pi^{3/2}} \, U_{\sigma,b}\left( Q_{u} \right) \, V_{\sigma,b}\left( Q_{v} \right),
		\end{aligned}
	\end{align}
	have infinite transverse power and are not square integrable with respect to transverse Lebesgue measure.
	We use the even and odd trigonometric functions $\mathcal{T}_{\mathrm{e}} \left( \cdot \right) = \cos \left( \cdot \right)$ and $\mathcal{T}_{\mathrm{o}} \left( \cdot \right) = \sin \left( \cdot \right)$, the Bessel function of the first kind $J_{\ell} \left( \cdot \right)$, the angular Mathieu function $Y_{\sigma,m} \left( \cdot \right)$, and the radial Mathieu function of the first kind $M_{\sigma,m} \left( \cdot \right)$~\cite{RodriguezLara2008p033813,DominguezCruz2026ae929a}.
	The Weber profiles contain the Euler gamma function $\Gamma \left( \cdot \right)$ and the real parabolic cylinder functions $U_{\sigma,b} \left( \cdot \right)$ and $V_{\sigma,b} \left( \cdot \right)$~\cite{RodriguezLara2010p327,DominguezCruz2026ae929a}, while the Kronecker delta $\delta_{ij}$ equals one when the indices match and zero otherwise.
	The mode labels,
	\begin{align}
		\begin{aligned}
			\lambda_{\mathrm{C}} =&~ \left( \bm{\sigma},\vartheta \right), \\
			\lambda_{\mathrm{B}} =&~ \left( \sigma,\ell \right), \\
			\lambda_{\mathrm{M}} =&~ \left( \sigma,m \right), \\
			\lambda_{\mathrm{W}} =&~ \left( \sigma,b \right),
		\end{aligned}
	\end{align}
	collect the Cartesian parity vector $\bm{\sigma} = (\sigma_{x},\sigma_{y})$, whose components $\sigma_{x},\sigma_{y} \in \{ \mathrm{e},\mathrm{o} \}$ identify even or odd profiles under the reflections $Q_{x} \mapsto - Q_{x}$ and $Q_{y} \mapsto - Q_{y}$, and the parity label $\sigma \in \{ \mathrm{e},\mathrm{o} \}$ under $Q_{y} \mapsto - Q_{y}$ for the other three families.
	They also include the direction angle $\vartheta \in (0,\pi / 2)$, integer Bessel order $\ell \geq 0$ for even profiles and $\ell \geq 1$ for odd profiles, integer Mathieu order $m \geq 0$ for even profiles and $m \geq 1$ for odd profiles, or continuous Weber separation parameter $b \in \mathbb{R}$.
	The separation eigenvalues,
	\begin{align}
		\begin{aligned}
			a_{\lambda_{\mathrm{C}}}^{\left( \mathrm{C} \right)} =&~ \frac{1}{4} \cos\left( 2\vartheta \right), \\
			a_{\lambda_{\mathrm{B}}}^{\left( \mathrm{B} \right)} =&~ \ell^{2}, \\
			a_{\lambda_{\mathrm{M}}}^{\left( \mathrm{M} \right)} =&~ a_{\sigma,m}\left( q \right), \\
			a_{\lambda_{\mathrm{W}}}^{\left( \mathrm{W} \right)} =&~ b,
		\end{aligned}
	\end{align}
	are independent of parity for the Cartesian, Bessel, and Weber families, while the Mathieu characteristic values generally differ between the two parity sectors.
	We use the ellipticity parameter $q = Q_{f}^{2} / 4$ and write the even and odd Mathieu characteristic values as $a_{\mathrm{e},m} \left( q \right) = a_{m} \left( q \right)$ and $a_{\mathrm{o},m} \left( q \right) = b_{m} \left( q \right)$.
	
	We use the orthogonal coordinate systems of our spectral-contraction analysis~\cite{DominguezCruz2026p2606.26068}, with circular coordinates satisfying $Q_{x} + i Q_{y} = Q_{\rho} e^{i Q_{\theta}}$, radial coordinate $Q_{\rho} \geq 0$ and azimuthal angle $Q_{\theta} \in [0,2\pi)$.
	The elliptic coordinates satisfy $Q_{x} + i Q_{y} = Q_{f} \cosh \left( Q_{\xi} + i Q_{\eta} \right)$, with radial coordinate $Q_{\xi} \geq 0$, angular coordinate $Q_{\eta} \in [0,2\pi)$, and half-interfocal distance $Q_{f} > 0$.
	The parabolic coordinates satisfy $Q_{x} + i Q_{y} = (Q_{u} + i Q_{v})^{2} / 2$, with coordinate ranges $Q_{u} \in \mathbb{R}$ and $Q_{v} \geq 0$.
	
	\begin{figure}[t]
		\centering
		\includegraphics[width=86mm]{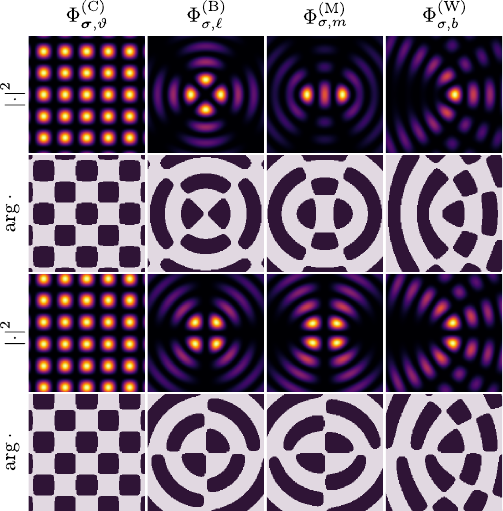}
		\caption{Columns show Cartesian waves with $\sigma_{x} = \mathrm{e}$ and $\vartheta = \pi / 4$, Bessel waves with $\ell = 2$, Mathieu waves with $m = 2$ and $q = 5$, and Weber waves with $b = 1$, from left to right.
			Rows show intensity and phase for even and odd parity on the range $Q_{x}, Q_{y} \in [ -10, 10]$.
			We normalize intensity to its maximum within the displayed window and use dark purple for zero phase and light gray for phase $\pi$.}
		\label{fig:Fig1}
	\end{figure}
	
	Figure~\ref{fig:Fig1} shows examples of the four Helmholtz families, all sharing reflection symmetry across the transverse $Q_{x}$ axis.
	The Cartesian profiles have a rectangular intensity grid whose nodal lines shift along $Q_{y}$ between parities.
	The Bessel profiles have concentric radial nodes and four angular lobes, with nodal diameters rotated by $\pi/4$ between parities.
	The Mathieu profiles have elliptic fringes and different central lobe arrangements, while the Weber profiles have curved parabolic fringes.
	The odd profiles share a nodal line along the transverse $Q_{x}$ axis, with opposite phases on either side despite their reflection-symmetric intensities.
	
	%%%%%%%%%%%%%%%%%%%%%%%
	\section{Angular spectra}
	\label{sec:Sec3}
	%%%%%%%%%%%%%%%%%%%%%%%
	
	The Whittaker angular-spectrum representation~\cite{Whittaker1903p333},
	\begin{align}
		\Phi\left( Q_{x},Q_{y} \right) = \frac{1}{\sqrt{2\pi}} \int_{0}^{2\pi} \alpha\left( \phi \right) e^{i\left( Q_{x} \cos \phi + Q_{y} \sin \phi \right)} \, \mathrm{d}\phi,
	\end{align}
	expresses the transverse profile $\Phi$ in terms of the plane-wave amplitude $\alpha$ over transverse propagation angles $\phi \in [0,2\pi)$.
	The generators of transverse translations and axial orbital angular momentum~\cite{Miller1974p626},
	\begin{align}
		\begin{aligned}
			\hat{P}_{x} =&~ \cos \phi, \\
			\hat{P}_{y} =&~ \sin \phi, \\
			\hat{L}_{z} =&~ - i \partial_{\phi},
		\end{aligned}
	\end{align}
	act by multiplication and differentiation on periodic angular amplitudes and close the Euclidean algebra $\mathrm{e}(2)$, with quadratic Casimir,
	\begin{align}
		\hat{P}^{2} = \cos^{2}\phi + \sin^{2}\phi = 1,
	\end{align}
	that places every constituent plane wave on the unit Helmholtz shell.
	
	The Cartesian, Bessel, Mathieu, and Weber separation operators~\cite{Boyer1976p35},
	\begin{align}
		\begin{aligned}
			\hat{A}_{\mathrm{C}} =&~ \frac{1}{4} \cos\left( 2\phi \right), \\
			\hat{A}_{\mathrm{B}} =&~ - \partial_{\phi}^{2}, \\
			\hat{A}_{\mathrm{M}} =&~ - \partial_{\phi}^{2} + 2 q \cos\left( 2\phi \right), \\
			\hat{A}_{\mathrm{W}} =&~ - i \left( \sin \phi \, \partial_{\phi} + \frac{1}{2} \cos \phi \right),
		\end{aligned}
	\end{align}
	satisfy the separation eigenvalue equation,
	\begin{align}
		\hat{A}_{\nu} \alpha_{\lambda_{\nu}}^{\left( \nu \right)} = a_{\lambda_{\nu}}^{\left( \nu \right)} \alpha_{\lambda_{\nu}}^{\left( \nu \right)},
	\end{align}
	with the same mode labels and separation eigenvalues as the real transverse profiles, whose angular spectra~\cite{Miller1977,Jauregui2005p033411,RodriguezLara2008p033813,RodriguezLara2010p327},
	\begin{align}
		\begin{aligned}
			\alpha_{\bm{\sigma},\vartheta}^{\left( \mathrm{C} \right)}\left( \phi \right) =&~ \frac{1}{2} \sum_{s,t = \pm 1} \left( - i s \right)^{\delta_{\sigma_{x},\mathrm{o}}} \left( - i t \right)^{\delta_{\sigma_{y},\mathrm{o}}} \delta_{2\pi}\left( \phi - \phi_{st} \right), \\
			\alpha_{\sigma,\ell}^{\left( \mathrm{B} \right)}\left( \phi \right) =&~ \frac{\left( - i \right)^{\ell}}{\sqrt{\pi\left( 1 + \delta_{\ell,0} \right)}} \mathcal{T}_{\sigma}\left( \ell \phi \right), \\
			\alpha_{\sigma,m}^{\left( \mathrm{M} \right)}\left( \phi \right) =&~ \frac{\left( - i \right)^{m}}{\sqrt{\pi}} Y_{\sigma,m}\left( \phi \right), \\
			\alpha_{\sigma,b}^{\left( \mathrm{W} \right)}\left( \phi \right) =&~ \frac{\left[ - i \operatorname{sgn}\left( \sin \phi \right) \right]^{\delta_{\sigma,\mathrm{o}}}}{\sqrt{4\pi \left\lvert \sin \phi \right\rvert}} e^{i b \ln\left\lvert \tan\left( \frac{\phi}{2} \right) \right\rvert},
		\end{aligned}
	\end{align}
	keep the even and odd reflection parities of their configuration-space counterparts.
	The Cartesian direction signs $s,t \in \{ -1,1 \}$ determine the four propagation angles $\phi_{st} = \arg \left( s \cos \left( \vartheta \right) + i t \sin \left( \vartheta \right) \right)$ modulo $2\pi$, where $\arg \left( \cdot \right)$ is the complex argument, $\delta_{2\pi} \left( \cdot \right)$ is the periodic Dirac delta, and $\operatorname{sgn} \left( \cdot \right)$ is the sign function.
	
	We use the angular inner product,
	\begin{align}
		\left\langle \alpha,\beta \right\rangle = \int_{0}^{2\pi} \alpha^{*}\left( \phi \right) \beta\left( \phi \right) \, \mathrm{d}\phi,
	\end{align}
	for angular amplitudes $\alpha$ and $\beta$, with the asterisk indicating complex conjugation, to obtain the normalization relations~\cite{Miller1977,Jauregui2005p033411,RodriguezLara2008p033813,RodriguezLara2010p327},
	\begin{align}
		\begin{aligned}
			\left\langle \alpha_{\bm{\sigma},\vartheta}^{\left( \mathrm{C} \right)},\alpha_{\bm{\sigma}^{\prime},\vartheta^{\prime}}^{\left( \mathrm{C} \right)} \right\rangle =&~ \delta_{\sigma_{x},\sigma^{\prime}_{x}} \delta_{\sigma_{y},\sigma^{\prime}_{y}} \delta\left( \vartheta - \vartheta^{\prime} \right), \\
			\left\langle \alpha_{\sigma,\ell}^{\left( \mathrm{B} \right)},\alpha_{\sigma^{\prime},\ell^{\prime}}^{\left( \mathrm{B} \right)} \right\rangle =&~ \delta_{\sigma,\sigma^{\prime}} \delta_{\ell,\ell^{\prime}}, \\
			\left\langle \alpha_{\sigma,m}^{\left( \mathrm{M} \right)},\alpha_{\sigma^{\prime},m^{\prime}}^{\left( \mathrm{M} \right)} \right\rangle =&~ \delta_{\sigma,\sigma^{\prime}} \delta_{m,m^{\prime}}, \\
			\left\langle \alpha_{\sigma,b}^{\left( \mathrm{W} \right)},\alpha_{\sigma^{\prime},b^{\prime}}^{\left( \mathrm{W} \right)} \right\rangle =&~ \delta_{\sigma,\sigma^{\prime}} \delta\left( b - b^{\prime} \right),
		\end{aligned}
	\end{align}
	where the primed labels identify the second mode, the Mathieu inner product uses the same ellipticity parameter $q$ for both modes, and $\delta \left( \cdot \right)$ is the Dirac delta.
	
	\begin{figure}[t]
		\centering
		\includegraphics[width=86mm]{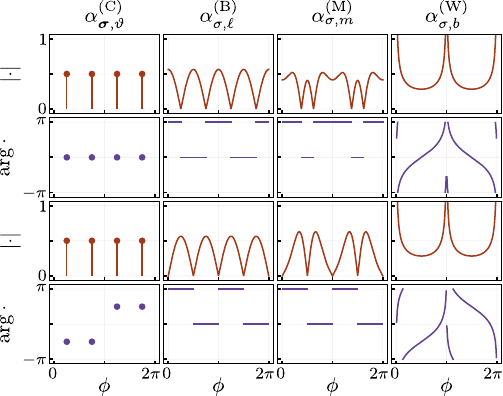}
		\caption{Columns show Cartesian waves with $\sigma_{x} = \mathrm{e}$ and $\vartheta = \pi / 4$, Bessel waves with $\ell = 2$, Mathieu waves with $m = 2$ and $q = 5$, and Weber waves with $b = 1$, from left to right.
			Rows show magnitude and phase for even and odd parity over $\phi\in[0,2\pi)$, using the stated spectral normalization.
			Cartesian stems and dots show delta-weight magnitudes and phases, while we clip the divergent Weber magnitudes at the upper limit of the common magnitude scale.}
		\label{fig:Fig2}
	\end{figure}
	
	Figure~\ref{fig:Fig2} shows angular spectra for the examples in Fig.~\ref{fig:Fig1}.
	The Cartesian spectra have four equal-magnitude delta weights, with phase zero for even parity and opposite $\pi/2$ phases on the two angular semicircles for odd parity.
	The Bessel spectra have equally spaced magnitude zeros shifted by $\pi/4$ between parities.
	The Mathieu spectra have different lobe shapes and zero locations for the two parities; both families have phases zero or $\pi$ between zeros.
	The Weber spectra have identical even and odd magnitudes that diverge at $\phi=0,\pi,2\pi$, with varying phases displayed modulo $2\pi$.
	Their odd phase differs from the even phase by opposite $\pi/2$ shifts on the two semicircles.
	
	%%%%%%%%%%%%%%%%%%%%%%%
	\section{Reflection grading}
	\label{sec:Sec4}
	%%%%%%%%%%%%%%%%%%%%%%%
	
	We define the reflection operator across the transverse $Q_{x}$ axis through its action in configuration space and angular-spectrum space,
	\begin{align}
		\begin{aligned}
			\hat{R} \Phi\left( Q_{x},Q_{y} \right) =&~ \Phi\left( Q_{x},- Q_{y} \right), \\
			\hat{R} \alpha\left( \phi \right) =&~ \alpha\left( - \phi \right),
		\end{aligned}
	\end{align}
	whose square is the identity,
	\begin{align}
		\hat{R}^{2} = 1,
	\end{align}
	with even and odd reflection sectors,
	\begin{align}
		\hat{R} \alpha_{\lambda_{\nu}}^{(\nu)} = \begin{cases}
			+ \alpha_{\lambda_{\nu}}^{(\nu)}, & \sigma = \mathrm{e}, \\
			- \alpha_{\lambda_{\nu}}^{(\nu)}, & \sigma = \mathrm{o},
		\end{cases}
	\end{align}
	where the reflection-parity label $\sigma$ is part of the mode label $\lambda_{\nu}$ and equals the $Q_{y}$-parity label $\sigma_{y}$ for Cartesian modes.
	
	From the reflection action on the Euclidean generators,
	\begin{align}
		\begin{aligned}
			\hat{R} \hat{P}_{x} \hat{R} =&~ \hat{P}_{x}, \\
			\hat{R} \hat{P}_{y} \hat{R} =&~ - \hat{P}_{y}, \\
			\hat{R} \hat{L}_{z} \hat{R} =&~ - \hat{L}_{z},
		\end{aligned}
	\end{align}
	we define our reflection $\mathbb{Z}_{2}$ grading through the relation,
	\begin{align}
		\hat{R} \hat{X} \hat{R} = \left( -1 \right)^{\left\lvert \hat{X} \right\rvert} \hat{X},
	\end{align}
	where the degree $\lvert\hat{X}\rvert\in\{0,1\}$ identifies an operator as even or odd,
	\begin{align}
		\begin{aligned}
			\left\lvert \hat{P}_{x} \right\rvert = \left\lvert \hat{R} \right\rvert =&~ 0, \\
			\left\lvert \hat{P}_{y} \right\rvert = \left\lvert \hat{L}_{z} \right\rvert =&~ 1.
		\end{aligned}
	\end{align}
	For operators $\hat{X}$ and $\hat{Y}$ of definite degree, our grading assigns the product degree,
	\begin{align}
		\left\lvert \hat{X} \hat{Y} \right\rvert = \left\lvert \hat{X} \right\rvert + \left\lvert \hat{Y} \right\rvert \bmod 2,
	\end{align}
	and gives even degree to the Helmholtz Casimir and separation operators,
	\begin{align}
		\left\lvert \hat{P}^{2} \right\rvert = \left\lvert \hat{A}_{\nu} \right\rvert = 0.
	\end{align}
	We decompose an operator without definite parity into its even and odd parts on the common domain,
	\begin{align}
		\begin{aligned}
			\hat{X}_{0} =&~ \frac{1}{2} \left( \hat{X} + \hat{R} \hat{X} \hat{R} \right), \\
			\hat{X}_{1} =&~ \frac{1}{2} \left( \hat{X} - \hat{R} \hat{X} \hat{R} \right).
		\end{aligned}
	\end{align}
	
	We use the standard supercommutator of Lie superalgebra theory~\cite{Kac1977p8},
	\begin{align}
		\left[ \hat{X}, \hat{Y} \right]_{\mathrm{s}} = \hat{X} \hat{Y} - \left( -1 \right)^{\left\lvert \hat{X} \right\rvert \left\lvert \hat{Y} \right\rvert} \hat{Y} \hat{X},
	\end{align}
	which reduces to an anticommutator when both operators are odd and a commutator otherwise.
	We include products beyond quadratic order in our construction because the even-operator commutator,
	\begin{align}
		\left[ \hat{L}_{z}^{2}, \hat{P}_{y}^{2} \right]_{\mathrm{s}} = - 2 i \left( \hat{L}_{z} \hat{P}_{x} \hat{P}_{y} + \hat{P}_{x} \hat{P}_{y} \hat{L}_{z} \right),
	\end{align}
	contains cubic products.
	We construct our superalgebra by adjoining reflection to the Euclidean enveloping algebra,
	\begin{align}
		\begin{aligned}
			\mathrm{s} =&~ \mathrm{s}_{0} \oplus \mathrm{s}_{1}, \\
			\mathrm{s}_{r} =&~ \operatorname{span}\left\{ \hat{P}_{x}^{j} \hat{P}_{y}^{k} \hat{L}_{z}^{\ell} \hat{R}^{t} \;\middle|\; k + \ell \equiv r \bmod 2 \right\},
		\end{aligned}
	\end{align}
	where $j,k,\ell$ are nonnegative integer powers, $r \in \{0,1\}$ labels the parity sector, $t \in \{0,1\}$ is the reflection exponent, and the operator products act on the unit shell.
	Our even and odd subspaces close under the supercommutator,
	\begin{align}
		\left[ \mathrm{s}_{r}, \mathrm{s}_{r^{\prime}} \right]_{\mathrm{s}} \subseteq \mathrm{s}_{r + r^{\prime} \bmod 2},
	\end{align}
	with parity sector label $r^{\prime} \in \{0,1\}$; arbitrary powers of the rotation generator make our superalgebra infinite-dimensional.
	The even Helmholtz Casimir commutes with every element of our superalgebra,
	\begin{align}
		\left[ \hat{P}^{2}, \hat{X} \right]_{\mathrm{s}} = 0,
	\end{align}
	for every operator $\hat{X} \in \mathrm{s}$ on the unit shell.
	
	%%%%%%%%%%%%%%%%%%%%%%%
	\section{Displaced OAM}
	\label{sec:Sec5}
	%%%%%%%%%%%%%%%%%%%%%%%
	
	Our expression of the Mathieu separation operator through OAM generators about the elliptic foci~\cite{RodriguezLara2008p033813},
	\begin{align}
		\hat{A}_{\mathrm{M}} = \frac{1}{2} \left[ \hat{L}_{z}+Q_{f}\hat{P}_{y},\hat{L}_{z}-Q_{f}\hat{P}_{y} \right]_{\mathrm{s}} + \frac{Q_{f}^{2}}{2},
	\end{align}
	motivates extending this construction to rotation axes along the reflection line,
	\begin{align}
		\begin{aligned}
			\hat{L}_{z, \eta} =&~ \hat{U}_{\eta} \hat{L}_{z} \hat{U}_{\eta}^{\dagger} \\
			=&~ \hat{L}_{z} + \eta \hat{P}_{y} \\
			=&~ - i \partial_{\phi} + \eta \sin \phi,
		\end{aligned}
	\end{align}
	with unitary translation operator,
	\begin{align}
		\hat{U}_{\eta} = e^{i\eta\hat{P}_{x}} = e^{i\eta\cos\phi}
	\end{align}
	which translates the rotation axis to $(Q_{x},Q_{y}) = (-\eta,0)$ for real dimensionless displacement $\eta$, and commutes with reflection $[ \hat{R}, \hat{U}_{\eta} ] =0$, preserving the odd parity of our displaced OAM generator.
	
	From our displaced OAM generators, we recover the three generators of the Euclidean algebra $\mathrm{e}(2)$,
	\begin{align}
		\begin{aligned}
			\hat{P}_{x} =&~ \frac{i}{\eta} \left[ \hat{L}_{z,0}, \hat{L}_{z,\eta} \right], \\
			\hat{P}_{y} =&~ \frac{1}{\eta} \left( \hat{L}_{z,\eta} - \hat{L}_{z,0} \right), \\
			\hat{L}_{z} =&~ \hat{L}_{z, 0},
		\end{aligned}
	\end{align}
	for any nonzero displacement $\eta$.
	From the supercommutator of two displaced OAM generators,
	\begin{align}
		\begin{aligned}
			\frac{1}{2} \left[ \hat{L}_{z,\eta},\hat{L}_{z,\eta^{\prime}} \right]_{\mathrm{s}} =&~ \hat{L}_{z}^{2} + \frac{\eta + \eta^{\prime}}{2} \left[ \hat{L}_{z}, \hat{P}_{y} \right]_{\mathrm{s}}  + \eta\eta^{\prime} \hat{P}_{y}^{2} \\
			=&~ \hat{A}_{\mathrm{B}} + \left( \eta + \eta^{\prime} \right) \hat{A}_{\mathrm{W}} + \eta\eta^{\prime} \left( \frac{1}{2} - 2 \hat{A}_{\mathrm{C}} \right),
		\end{aligned}
	\end{align}
	with real displacements $\eta$ and $\eta^{\prime}$, we recover the four separation operators,
	\begin{align}
		\begin{aligned}
			\hat{A}_{\mathrm{C}} =&~  \frac{1}{4} \lim_{\eta\to\infty} \left( 1 +  \frac{1}{\eta^{2}} \left[ \hat{L}_{z,\eta},\hat{L}_{z,-\eta} \right]_{\mathrm{s}} \right), \\
			\hat{A}_{\mathrm{B}} =&~ \frac{1}{2} \left[ \hat{L}_{z,0},\hat{L}_{z,0} \right]_{\mathrm{s}}, \\
			\hat{A}_{\mathrm{M}} =&~  \frac{1}{2} \left( Q_{f}^{2} +  \left[ \hat{L}_{z,Q_{f}},\hat{L}_{z,-Q_{f}} \right]_{\mathrm{s}} \right),\\
			\hat{A}_{\mathrm{W}} =&~ \frac{1}{2} \lim_{\eta\to\infty}  \frac{1}{\eta} \left[ \hat{L}_{z,0},\hat{L}_{z,\eta} \right]_{\mathrm{s}},
		\end{aligned}
	\end{align}
	by choosing the placement of the rotation axes.
	The Cartesian limit places the axes at opposite infinities, while Bessel separation places both axes at the origin.
	Mathieu separation places the axes at the elliptic foci~\cite{Erikson1949p29,Coulson1967p337,RodriguezLara2008p033813}, while the Weber limit keeps one axis at the origin and sends the other to infinity~\cite{Miller1974p626,RodriguezLara2009055806,Bandres2013p013054}.
	These limits agree with the contractions connecting the corresponding coordinate systems~\cite{Boyer1976p35,Izmestev1996p5949}.
	
	%%%%%%%%%%%%%%%%%%%%%%%
	\section{Witten supersymmetry}
	\label{sec:Sec6}
	%%%%%%%%%%%%%%%%%%%%%%%
	
	A $\mathbb{Z}_{2}$ grading can support a Witten supersymmetric model when the grading operator anticommutes with a self-adjoint operator~\cite{Witten1982p253};
	for example, the anticommutation between the reflection operator $\hat{R}$ and the self-adjoint OAM generator $\hat{L}_{z}$,
	\begin{align}
		\hat{R} \hat{L}_{z} + \hat{L}_{z}  \hat{R} = 0,
	\end{align}
	realizes the known hidden supersymmetry of a quantum free particle on a circle~\cite{Correa2007p2493}.
	
	We extend this construction to our self-adjoint displaced OAM generators,
	\begin{align}
		\hat{R} \hat{L}_{z,\eta} + \hat{L}_{z,\eta} \hat{R} = 0,
	\end{align}
	for real displacement $\eta$, using the reflection-invariant domain of the OAM generator.
	We construct our model defining the reflection projectors,
	\begin{align}
		\hat{\Pi}_{\pm} = \frac{1}{2} \left( 1 \pm \hat{R} \right),
	\end{align}
	which select the even and odd reflection sectors in configuration or angular-spectrum spaces.
	Using the projector relations for displaced OAM,
	\begin{align}
		\hat{L}_{z,\eta} \hat{\Pi}_{\pm} = \hat{\Pi}_{\mp} \hat{L}_{z,\eta},
	\end{align}
	we define our non-Hermitian charges,
	\begin{align}
		\begin{aligned}
			\hat{\mathcal{Q}}_{\eta} =&~ \hat{\Pi}_{-} \hat{L}_{z,\eta} = \hat{L}_{z,\eta} \hat{\Pi}_{+}, \\
			\hat{\mathcal{Q}}_{\eta}^{\dagger} =&~ \hat{\Pi}_{+} \hat{L}_{z,\eta} = \hat{L}_{z,\eta} \hat{\Pi}_{-},
		\end{aligned}
	\end{align}
	with $\hat{\mathcal{Q}}_{\eta}$ ($\hat{\mathcal{Q}}_{\eta}^{\dagger}$) mapping even (odd) reflection states to odd (even) reflection states.
	We decompose our displaced OAM generator into nilpotent parts,
	\begin{align}
		\begin{aligned}
			\hat{L}_{z,\eta} =&~ \hat{\mathcal{Q}}_{\eta} + \hat{\mathcal{Q}}_{\eta}^{\dagger}, \\
			\hat{\mathcal{Q}}_{\eta}^{2} =&~ \hat{\mathcal{Q}}_{\eta}^{\dagger 2} = 0,
		\end{aligned}
	\end{align}
	and define our super-Hamiltonian through their super-commutator,
	\begin{align}
		\hat{\mathcal{H}}_{\eta} = \left[ \hat{\mathcal{Q}}_{\eta},\hat{\mathcal{Q}}_{\eta}^{\dagger} \right]_{\mathrm{s}} = \hat{L}_{z,\eta}^{2},
	\end{align}
	with nonnegative spectrum.
	When the displaced OAM generator also commutes with a separation operator, we obtain an opposite-parity eigenmode at the same separation eigenvalue whenever the charge action is nonzero.
	
	For Cartesian separation, the rescaled OAM generator $\hat{L}_{z,\eta}/\eta$ approaches the odd translation generator $\hat{P}_{y}$ as the displacement $\eta\to\infty$.
	We recover the Cartesian charges through the same rescaling,
	\begin{align}
		\begin{aligned}
			\hat{\mathcal{Q}}_{\mathrm{C}} =&~ \lim_{\eta\to\infty} \frac{1}{\eta} \hat{\mathcal{Q}}_{\eta} = \hat{\Pi}_{-} \hat{P}_{y} = \hat{P}_{y} \hat{\Pi}_{+}, \\
			\hat{\mathcal{Q}}_{\mathrm{C}}^{\dagger} =&~ \lim_{\eta\to\infty} \frac{1}{\eta} \hat{\mathcal{Q}}_{\eta}^{\dagger} = \hat{\Pi}_{+} \hat{P}_{y} = \hat{P}_{y} \hat{\Pi}_{-},
		\end{aligned}
	\end{align}
	and obtain our super-Hamiltonian,
	\begin{align}
		\hat{\mathcal{H}}_{\mathrm{C}} = \left[ \hat{\mathcal{Q}}_{\mathrm{C}},\hat{\mathcal{Q}}_{\mathrm{C}}^{\dagger} \right]_{\mathrm{s}} = \hat{P}_{y}^{2} = \frac{1}{2} - 2 \hat{A}_{\mathrm{C}},
	\end{align}
	using the commutation of $\hat{P}_{y}$ with the Cartesian separator to preserve its eigenvalues in our charge actions.
	The generalized Cartesian eigenmodes of our super-Hamiltonian,
	\begin{align}
		\hat{\mathcal{H}}_{\mathrm{C}} \alpha_{\bm{\sigma},\vartheta}^{\left( \mathrm{C} \right)} = \sin^{2}\vartheta \, \alpha_{\bm{\sigma},\vartheta}^{\left( \mathrm{C} \right)},
	\end{align}
	form opposite-parity pairs at the same separation eigenvalue for either $Q_{x}$-parity label $\sigma_{x}$ and $0<\vartheta<\pi/2$ under our charge actions,
	\begin{align}
		\begin{aligned}
			\hat{\mathcal{Q}}_{\mathrm{C}} \alpha_{\left( \sigma_{x},\mathrm{e} \right),\vartheta}^{\left( \mathrm{C} \right)} =&~ i \sin\vartheta \, \alpha_{\left( \sigma_{x},\mathrm{o} \right),\vartheta}^{\left( \mathrm{C} \right)}, \\
			\hat{\mathcal{Q}}_{\mathrm{C}}^{\dagger} \alpha_{\left( \sigma_{x},\mathrm{o} \right),\vartheta}^{\left( \mathrm{C} \right)} =&~ - i \sin\vartheta \, \alpha_{\left( \sigma_{x},\mathrm{e} \right),\vartheta}^{\left( \mathrm{C} \right)}, \\
			\hat{\mathcal{Q}}_{\mathrm{C}} \alpha_{\left( \sigma_{x},\mathrm{o} \right),\vartheta}^{\left( \mathrm{C} \right)} =&~ \hat{\mathcal{Q}}_{\mathrm{C}}^{\dagger} \alpha_{\left( \sigma_{x},\mathrm{e} \right),\vartheta}^{\left( \mathrm{C} \right)} = 0.
		\end{aligned}
	\end{align}
	The super-Hamiltonian has no zero mode with a square-integrable angular amplitude; at direction angle $\vartheta=0$, outside our Cartesian label range, the odd $Q_{y}$-parity profiles vanish, leaving generalized even zero modes without odd Cartesian partners.
	We cannot assign a Fredholm Witten index to this realization because zero lies in the continuous spectrum of our super-Hamiltonian~\cite{Carey2016p881}.
	
	For Bessel separation, we take $\eta=0$ and use the OAM generator $\hat{L}_{z,0}=\hat{L}_{z}$.
	We recover our Bessel charges,
	\begin{align}
		\begin{aligned}
			\hat{\mathcal{Q}}_{\mathrm{B}} =&~ \hat{\mathcal{Q}}_{0} = \hat{\Pi}_{-} \hat{L}_{z} = \hat{L}_{z} \hat{\Pi}_{+}, \\
			\hat{\mathcal{Q}}_{\mathrm{B}}^{\dagger} =&~ \hat{\mathcal{Q}}_{0}^{\dagger} = \hat{\Pi}_{+} \hat{L}_{z} = \hat{L}_{z} \hat{\Pi}_{-},
		\end{aligned}
	\end{align}
	and obtain our super-Hamiltonian,
	\begin{align}
		\hat{\mathcal{H}}_{\mathrm{B}} = \left[ \hat{\mathcal{Q}}_{\mathrm{B}},\hat{\mathcal{Q}}_{\mathrm{B}}^{\dagger} \right]_{\mathrm{s}} = \hat{L}_{z}^{2} = \hat{A}_{\mathrm{B}},
	\end{align}
	which coincides with the Bessel separation operator.
	The Bessel eigenmodes of our super-Hamiltonian,
	\begin{align}
		\hat{\mathcal{H}}_{\mathrm{B}} \alpha_{\sigma,\ell}^{\left( \mathrm{B} \right)} = \ell^{2} \alpha_{\sigma,\ell}^{\left( \mathrm{B} \right)},
	\end{align}
	form opposite-parity pairs at each order $\ell\geq1$ under our charge actions,
	\begin{align}
		\begin{aligned}
			\hat{\mathcal{Q}}_{\mathrm{B}} \alpha_{\mathrm{e},\ell}^{\left( \mathrm{B} \right)} =&~ i \ell \alpha_{\mathrm{o},\ell}^{\left( \mathrm{B} \right)}, \\
			\hat{\mathcal{Q}}_{\mathrm{B}}^{\dagger} \alpha_{\mathrm{o},\ell}^{\left( \mathrm{B} \right)} =&~ - i \ell \alpha_{\mathrm{e},\ell}^{\left( \mathrm{B} \right)}, \\
			\hat{\mathcal{Q}}_{\mathrm{B}} \alpha_{\mathrm{o},\ell}^{\left( \mathrm{B} \right)} =&~ \hat{\mathcal{Q}}_{\mathrm{B}}^{\dagger} \alpha_{\mathrm{e},\ell}^{\left( \mathrm{B} \right)} = 0.
		\end{aligned}
	\end{align}
	Both charges annihilate the even zero mode $\alpha_{\mathrm{e},0}^{(\mathrm{B})} \left( \phi \right) = 1 / \sqrt{2\pi}$, whose lack of an odd partner gives the Witten index~\cite{Witten1982p253},
	\begin{align}
		\Delta = n_{\mathrm{e}}^{(0)} - n_{\mathrm{o}}^{(0)} = 1,
	\end{align}
	where $n_{\mathrm{e}}^{(0)} = 1$ and $n_{\mathrm{o}}^{(0)} = 0$ count zero modes in each parity sector.
	The supersymmetry is unbroken in both configuration and angular-spectrum spaces.
	Since the unitary translation operator $\hat{U}_{\eta}$ commutes with reflection, we obtain a unitarily equivalent copy of our unbroken Bessel supersymmetry model at every real displacement $\eta$, with even zero mode $\alpha_{\eta,0} \left( \phi \right) = e^{i\eta\cos \left( \phi \right)} / \sqrt{2\pi}$, corresponding to configuration-space profile $J_{0} \left( \sqrt{\left( Q_{x} + \eta \right)^{2} + Q_{y}^{2}} \right)$ centered at $(-\eta,0)$, and Witten index $\Delta=1$.

	For Mathieu separation, we choose the focal displacements $\eta=\pm Q_{f}$ and obtain super-Hamiltonians $\hat{\mathcal{H}}_{\pm Q_{f}}=\hat{L}_{z,\pm Q_{f}}^{2}$ from our charges, but the focal generators do not commute with the Mathieu separator,
	\begin{align}
		\begin{aligned}
			\left[ \hat{A}_{\mathrm{M}},\hat{L}_{z,\pm Q_{f}} \right] =&~ \mp i Q_{f} \left( \hat{P}_{x} \hat{L}_{z,\pm Q_{f}} + \hat{L}_{z,\pm Q_{f}} \hat{P}_{x} \right), \\
			=&~ -4iq\sin 2\phi \pm Q_{f} \left(\sin \phi - 2\cos\phi \, \partial_{\phi} \right).
		\end{aligned}
	\end{align}
	For a Mathieu eigenmode, we evaluate whether the focal action preserves its separation eigenvalue,
	\begin{align}
		\left( \hat{A}_{\mathrm{M}}-a_{\sigma,m}\left( q \right) \right)\hat{L}_{z,\pm Q_{f}}\alpha_{\sigma,m}^{(\mathrm{M})}
		=\left[ \hat{A}_{\mathrm{M}},\hat{L}_{z,\pm Q_{f}} \right]\alpha_{\sigma,m}^{(\mathrm{M})},
	\end{align}
	with the parity-reversed field generally outside the eigenspace of the original Mathieu separation eigenvalue.
	At $q=0$, we recover the Bessel pairing, while in the large-focal-distance limit we recover the Cartesian generators and charges after rescaling.
	From the simple periodic spectrum $a_{\mathrm{e},0} \left( q \right)<a_{\mathrm{o},1} \left( q \right)<a_{\mathrm{e},1} \left( q \right)<a_{\mathrm{o},2} \left( q \right)<a_{\mathrm{e},2} \left( q \right)<\cdots$ for $q>0$~\cite{NIST2026Mathieu}, we exclude an alternative odd charge that preserves the Mathieu eigenvalues.
	An odd commuting operator must map each eigenstate into its one-dimensional eigenspace with opposite reflection parity, forcing its image to vanish.
	Such a closed operator is zero if its domain contains the complete Mathieu eigenbasis and its eigenstate images belong to the separator domain.
	
	For Weber separation, we test our displaced OAM generators through their commutator with the separator,
	\begin{align}
		\left[ \hat{A}_{\mathrm{W}}, \hat{L}_{z,\eta} \right] = \frac{i}{2} \left( \hat{L}_{z} \hat{P}_{x} + \hat{P}_{x} \hat{L}_{z} \right) - i \eta \hat{P}_{x} \hat{P}_{y},
	\end{align}
	which is nonzero for every constant displacement $\eta$.
	The rescaled translation generator also fails to preserve the Weber separation eigenvalue,
	\begin{align}
		\left[ \hat{A}_{\mathrm{W}}, \hat{P}_{y} \right] = - i \hat{P}_{x} \hat{P}_{y},
	\end{align}
	leading us to introduce the bounded self-adjoint sign operator,
	\begin{align}
		\begin{aligned}
			\hat{S} =&~ \operatorname{sgn}\left( \hat{P}_{y} \right), \\
			\hat{S}\alpha\left( \phi \right) =&~ \operatorname{sgn}\left( \sin\phi \right)\alpha\left( \phi \right),
		\end{aligned}
	\end{align}
	which anticommutes with reflection, commutes with the Weber separator, and satisfies $\hat{S}^{2}=1$.
	Using the self-adjoint odd product $\hat{S}\hat{A}_{\mathrm{W}}$, we construct our Weber charges,
	\begin{align}
		\begin{aligned}
			\hat{\mathcal{Q}}_{\mathrm{W}} =&~ \hat{\Pi}_{-} \hat{S} \hat{A}_{\mathrm{W}} = \hat{S} \hat{A}_{\mathrm{W}} \hat{\Pi}_{+}, \\
			\hat{\mathcal{Q}}_{\mathrm{W}}^{\dagger} =&~ \hat{\Pi}_{+} \hat{S} \hat{A}_{\mathrm{W}} = \hat{S} \hat{A}_{\mathrm{W}} \hat{\Pi}_{-},
		\end{aligned}
	\end{align}
	and their super-Hamiltonian,
	\begin{align}
		\hat{\mathcal{H}}_{\mathrm{W}} = \left[ \hat{\mathcal{Q}}_{\mathrm{W}},\hat{\mathcal{Q}}_{\mathrm{W}}^{\dagger} \right]_{\mathrm{s}} = \hat{A}_{\mathrm{W}}^{2},
	\end{align}
	with squared Weber separation eigenvalues.
	The generalized Weber eigenmodes of our super-Hamiltonian,
	\begin{align}
		\hat{\mathcal{H}}_{\mathrm{W}} \alpha_{\sigma,b}^{\left( \mathrm{W} \right)} = b^{2} \alpha_{\sigma,b}^{\left( \mathrm{W} \right)},
	\end{align}
	form opposite-parity pairs at the same separation eigenvalue for $b\neq0$ under our charge actions,
	\begin{align}
		\begin{aligned}
			\hat{\mathcal{Q}}_{\mathrm{W}} \alpha_{\mathrm{e},b}^{\left( \mathrm{W} \right)} =&~ i b \, \alpha_{\mathrm{o},b}^{\left( \mathrm{W} \right)}, \\
			\hat{\mathcal{Q}}_{\mathrm{W}}^{\dagger} \alpha_{\mathrm{o},b}^{\left( \mathrm{W} \right)} =&~ - i b \, \alpha_{\mathrm{e},b}^{\left( \mathrm{W} \right)}, \\
			\hat{\mathcal{Q}}_{\mathrm{W}} \alpha_{\mathrm{o},b}^{\left( \mathrm{W} \right)} =&~ \hat{\mathcal{Q}}_{\mathrm{W}}^{\dagger} \alpha_{\mathrm{e},b}^{\left( \mathrm{W} \right)} = 0,
		\end{aligned}
	\end{align}
	using the phases of the Weber spectra~\cite{RodriguezLara2010p327}.
	At separation parameter $b=0$, both charges annihilate the two generalized zero modes, while the sign operator $\hat{S}$ connects their opposite parities; neither mode is normalizable in angular space.
	We cannot assign a Fredholm Witten index to this realization because zero lies in the continuous spectrum of our Weber super-Hamiltonian~\cite{Carey2016p881}.
	We use the spectral sign of the Euclidean generator $\hat{P}_{y}$ in the odd factor $\hat{S}\hat{A}_{\mathrm{W}}$, extending our construction from polynomials in the generators to nonpolynomial functions of them.
	
	%%%%%%%%%%%%%%%%%%%%%%%
	\section{Supersymmetric Talbot sequences}
	\label{sec:Sec7}
	%%%%%%%%%%%%%%%%%%%%%%%
	
	We extend our unit-shell model to monochromatic, $k = 2 \pi / \lambda$, transverse shells with
	\begin{align}
		k_{\perp}^{(j)} = \sqrt{j} k_{\perp} < k
	\end{align}
	for constructing self-imaging Bessel waves in the paraxial regime~\cite{Montgomery1967p772,Hebri2019p4355}.
	On the direct sum of shell spaces, we define the multishell charges,
	\begin{align}
		\hat{\mathcal{Q}}_{\mathrm{multi}} = \bigoplus_{j} \hat{\mathcal{Q}}^{(j)},
	\end{align}
	with nilpotent, non-Hermitian supercharge and reflection projectors,
	\begin{align}
		\begin{aligned}
			\hat{\mathcal{Q}}^{(j)} =&~ \hat{\Pi}_{-}^{(j)} \hat{L}_{z, \eta_{j}}^{(j)}, \\
			\hat{\Pi}_{\pm}^{(j)} =&~ \frac{1}{2} \left( 1 \pm \hat{R}^{(j)} \right),
		\end{aligned}
	\end{align}
	acting in the $j$-th shell through the scaled coordinate, translation, OAM, and displaced OAM operators,
	\begin{align}
		\begin{aligned}
			\hat{Q}_{c}^{(j)} =&~ \sqrt{j}\hat{Q}_{c}, \\
			\hat{P}_{c}^{(j)} =&~ \frac{1}{\sqrt{j}}\hat{P}_{c}, \\
			\hat{L}_{z}^{(j)} =&~ \hat{Q}_{x}^{(j)}\hat{P}_{y}^{(j)}
			- \hat{Q}_{y}^{(j)}\hat{P}_{x}^{(j)}, \\
			\hat{L}_{z,\eta_j}^{(j)} =&~ \hat{L}_{z}^{(j)} + \eta_j\hat{P}_{y}^{(j)},
		\end{aligned}
	\end{align}
	with $c \in \{ x, y\}$ and $\eta_{j} = \sqrt{j} \eta$ for the common rotation axis at $(x, y) = ( - \eta / k_{\perp}, 0)$, and define the super-Hamiltonian,
	\begin{align}
		\hat{\mathcal{H}}_{\mathrm{multi}} = \bigoplus_{j} \left[\hat{\mathcal{Q}}^{(j)},\hat{\mathcal{Q}}^{(j)\dagger}\right]_{\mathrm{s}} = \bigoplus_{j} \left(\hat{L}_{z,\eta_j}^{(j)}\right)^{2},
	\end{align}
	providing a Witten superalgebra with index one on each shell and total index equal to the number of included shells.
	
	Exact forward propagation in the $j$-th shell,
	\begin{align}
		\Phi^{(j)}(z) = e^{i k_{z}^{(j)} z} \Phi^{(j)}(0),
	\end{align}
	with $k_{z}^{(j)} = \sqrt{k^{2} - j k_{\perp}^{2}}$, reduces in the paraxial regime $j k_{\perp}^{2} \ll k^{2}$ to propagation with equally spaced longitudinal wave numbers,
	\begin{align}
		k_{z}^{(j)} \simeq k - \frac{j k_{\perp}^{2}}{2k},
	\end{align}
	from which we construct the multishell field,
	\begin{align}
		\Phi_{\mathrm{multi}}(z)
		\simeq e^{ikz} \sum_{j}
		e^{- \frac{1}{2 k} i j k_{\perp}^{2} z} \Phi^{(j)}(0),
	\end{align}
	with Talbot recurrence,
	\begin{align}
		\begin{aligned}
			\Phi_{\mathrm{multi}}(z + z_{\mathrm{T}})
			=&~ e^{ikz_{\mathrm{T}}}\Phi_{\mathrm{multi}}(z), \\
			z_{\mathrm{T}} =&~ \frac{4\pi k}{k_{\perp}^{2}},
		\end{aligned}
	\end{align}
	within the paraxial approximation.
	Since our charges preserve each shell and commute with propagation,
	\begin{align}
		\left[ \hat{\mathcal{Q}}^{(j)}, e^{i k_{z}^{(j)} z} \right] = \left[ \hat{\mathcal{Q}}^{(j)\dagger}, e^{i k_{z}^{(j)} z} \right] = 0,
	\end{align}
	we obtain the same Talbot recurrence for every nonzero field generated by successive charge actions, independently of the initial reflection parity.
	
	From the Talbot field $\Phi_{\mathrm{multi}}$, we construct four initial partners,
	\begin{align}
		\begin{aligned}
			\Phi_{\mathrm{o}} =&~ \hat{\mathcal{Q}}_{\mathrm{multi}}\Phi_{\mathrm{multi}}, \\
			\Phi_{\mathrm{e}} =&~ \hat{\mathcal{Q}}_{\mathrm{multi}}^{\dagger}\Phi_{\mathrm{multi}}, \\
			\Phi_{1} =&~ \hat{\mathcal{Q}}_{1}\Phi_{\mathrm{multi}}
			= \Phi_{\mathrm{o}} + \Phi_{\mathrm{e}}, \\
			\Phi_{2} =&~ \hat{\mathcal{Q}}_{2}\Phi_{\mathrm{multi}}
			= i \left( \Phi_{\mathrm{e}} - \Phi_{\mathrm{o}} \right),
		\end{aligned}
	\end{align}
	from the non-Hermitian and the Hermitian charges,
	\begin{align}
		\begin{aligned}
			\hat{\mathcal{Q}}_{1} =&~
			\hat{\mathcal{Q}}_{\mathrm{multi}}
			+ \hat{\mathcal{Q}}_{\mathrm{multi}}^{\dagger}, \\
			\hat{\mathcal{Q}}_{2} =&~
			i \left( \hat{\mathcal{Q}}_{\mathrm{multi}}^{\dagger}
			- \hat{\mathcal{Q}}_{\mathrm{multi}} \right).
		\end{aligned}
	\end{align}
	The non-Hermitian charges give the odd (even) partner from the even (odd) component of the seed and the Hermitian charges combine both outputs with different relative phases.
	The two Hermitian-charge partners have reflected intensity patterns and phases related by $\Phi_{2} = i \hat{R} \Phi_{1}$.
	We continue the first two sequences by alternating $\hat{\mathcal{Q}}_{\mathrm{multi}}$ and $\hat{\mathcal{Q}}_{\mathrm{multi}}^{\dagger}$, and the other two by repeatedly applying the corresponding Hermitian charge.
	Since $\hat{\mathcal{Q}}_{1}^{2}=\hat{\mathcal{Q}}_{2}^{2}=\hat{\mathcal{H}}_{\mathrm{multi}}$, their even iterates coincide, while their odd iterates combine the two parity branches.
	For a finite Bessel superposition, we can continue indefinitely within the space of occupied OAM orders, changing their relative weights without changing the Talbot recurrence.
	
	For our ideal example, we prepare each shell in configuration and angular-spectrum spaces,
	\begin{align}
		\begin{aligned}
			\Phi^{(j)} \left( z \right) \simeq&~ \frac{1}{\sqrt{2}} e^{i k z} e^{- \frac{1}{2 k} i j k_{\perp}^{2} z} \\
			&\times \sum_{\ell, \sigma} c_{\sigma,\ell}^{(j)} \Phi_{\sigma,\ell}^{\left( \mathrm{B} \right)} \left( Q_{\rho,\eta_{j}}^{(j)}, Q_{\theta,\eta_{j}}^{(j)} \right), \\
			\alpha^{(j)} \left( \phi, z \right) \simeq&~ \frac{1}{\sqrt{2}} e^{i k z} e^{- \frac{1}{2 k} i j k_{\perp}^{2} z} e^{i \eta_{j} \cos \phi} \\
			&\times \sum_{\ell, \sigma} c_{\sigma,\ell}^{(j)} \alpha_{\sigma,\ell}^{\left( \mathrm{B} \right)} \left( \phi \right),
		\end{aligned}
	\end{align}
	with displaced circular coordinates,
	\begin{align}
		Q_{\rho,\eta_{j}}^{(j)} e^{i Q_{\theta,\eta_{j}}^{(j)}} = Q_{x}^{(j)} + \eta_{j} + i Q_{y}^{(j)},
	\end{align}
	and $\sigma \in \{ \mathrm{e}, \mathrm{o} \}$, then construct the multishell seed in both representations,
	\begin{align}
		\begin{aligned}
			\Phi_{\mathrm{multi}} \left( z \right) =&~ \sum_{j} \Phi^{(j)} \left( z \right), \\
			\alpha_{\mathrm{multi}} \left( z \right) =&~ \bigoplus_{j} \alpha^{(j)} \left( \phi, z \right),
		\end{aligned}
	\end{align}
	combining the configuration-space fields at common physical coordinates.
	We choose the pairs,
	\begin{align}
		\left( j, \ell \right) \in \left\{ \left( 1, 1 \right), \left( 2, 2 \right), \ldots, \left( 12, 12 \right) \right\},
	\end{align}
	with even reflection parity for odd $j$ and odd reflection parity for even $j$, assigning $c_{\sigma,j}^{(j)} = c_{j - 1}$ for the selected parity and zero to the other coefficients,
	\begin{align}
		\begin{aligned}
			\bm{c} =&~ \left( c_{0}, \ldots, c_{11} \right), \\
			c_{j - 1} \left( \beta \right) =&~ \left( \frac{\beta^{j - 1} / \left( j - 1 \right)!}{\sum_{r = 0}^{11} \beta^{r} / r!} \right)^{1/2},
		\end{aligned}
	\end{align}
	whose squared amplitudes follow a Poisson distribution with parameter $\beta > 0$, which we condition on $0 \leq j - 1 \leq 11$.
	For our example, we set $\beta = 5$, $k_{\perp} = 2 \pi / (0.5\,\mathrm{mm})$, $\lambda = 633$ nm, and $\eta / k_{\perp} = 0.1$ mm.
	These parameters give a Talbot distance $z_{\mathrm{T}} = 789.8894$ mm and an eighth-period spacing of $98.7362$ mm.
	We displace the common rotation axis to $(-0.1, 0)$ mm and center the displayed window at the laboratory origin.
	
	\begin{figure*}[t]
		\centering
		\includegraphics[width=\textwidth]{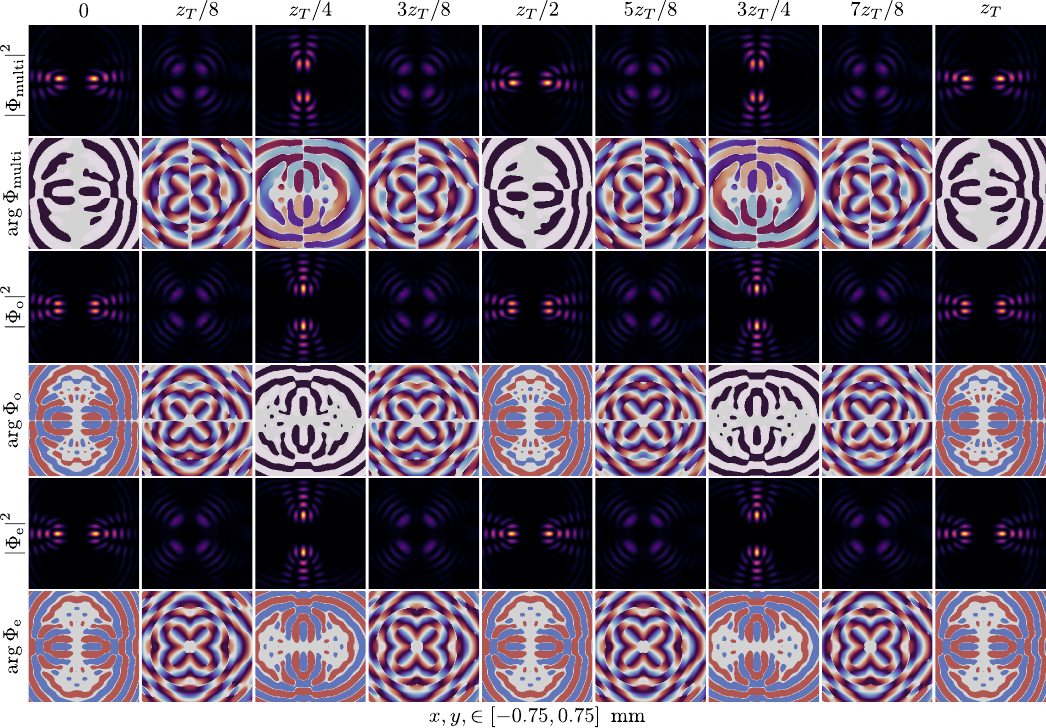}
		\caption{Ideal Talbot seed and its two non-Hermitian-charge partners.
			Successive row pairs show intensity and phase of $\Phi_{\mathrm{multi}}$, $\Phi_{\mathrm{o}}$, and $\Phi_{\mathrm{e}}$.
			Columns show $z / z_{\mathrm{T}} = 0, 1/8, 1/4, 3/8, 1/2, 5/8, 3/4, 7/8, 1$.
			All panels span $x, y \in [ -0.75, 0.75 ]$ mm.
			Each field uses one intensity maximum across all nine planes, and phase spans $[ -\pi, \pi ]$. We mask regions with intensity below $10^{-3}$ of that maximum in light gray.}
		\label{fig:Fig3}
	\end{figure*}
	
	In Fig.~\ref{fig:Fig3}, we show the intensity and phase of the seed and its non-Hermitian-charge partners in successive row pairs, at propagation distances $z / z_{\mathrm{T}} = 0, 1/8, 1/4, 3/8, 1/2, 5/8, 3/4, 7/8, 1$ from left to right.
	The seed has two dominant groups of intensity maxima arranged horizontally at the initial plane and vertically at the quarter-period plane, with weaker surrounding fringes.
	Through the charges, we weight each occupied OAM order by $\ell$, increasing the relative contribution of higher orders and changing the central lobes and surrounding fringes.
	The odd partner changes sign across the horizontal axis, while the even partner retains its value under reflection.
	All three fields recover their initial intensity and carrier-removed phase at the full Talbot period.
	
	%%%%%%%%%%%%%%%%%%%%%%%
	\section{Experimental Talbot sequences}
	\label{sec:Sec8}
	%%%%%%%%%%%%%%%%%%%%%%%
	
	For finite power, we apply the same physical Gaussian to each shell field,
	\begin{align}
		\begin{aligned}
			G^{(j)} =&~ e^{- \frac{1}{ w_{\mathrm{G}}^{(j) 2}} \left[ \left( Q_{x}^{(j)} + \eta_{j} \right)^{2} + \left( Q_{y}^{(j)} \right)^{2} \right]}, \\
			w_{\mathrm{G}}^{(j)} =&~ k_{\perp}^{(j)} w_{0},
		\end{aligned}
	\end{align}
	with common amplitude radius $w_{0}$.
	Apodization broadens each ring, with the shell index $j$ now identifying its input contribution.
	We propagate each complete input spectrum through the forward angular-spectrum operator,
	\begin{align}
		\Phi_{\mathrm{G}}^{(j)} \left( z \right) = e^{i z \sqrt{k^{2} - \left( k_{\perp}^{(j)} \right)^{2} \hat{P}^{(j) 2}}} G^{(j)} \Phi^{(j)} \left( 0 \right),
	\end{align}
	with dimensionless squared transverse-wavevector operator $\hat{P}^{(j) 2} = \hat{P}_{x}^{(j) 2} + \hat{P}_{y}^{(j) 2}$ on the propagating spectral subspace, and sum the contributions at common physical coordinates.
	With the Gaussian factor centered on the common rotation axis, both charges commute with Gaussian apodization,
	\begin{align}
		\left[ \hat{\mathcal{Q}}^{(j)}, G^{(j)} \right] = \left[ \hat{\mathcal{Q}}^{(j)\dagger}, G^{(j)} \right] = 0,
	\end{align}
	and with the exact forward-propagation operator,
	\begin{align}
		\begin{aligned}
			\left[ \hat{\mathcal{Q}}^{(j)}, e^{i z \sqrt{k^{2} - \left( k_{\perp}^{(j)} \right)^{2} \hat{P}^{(j) 2}}} \right] =&~ 0, \\
			\left[ \hat{\mathcal{Q}}^{(j)\dagger}, e^{i z \sqrt{k^{2} - \left( k_{\perp}^{(j)} \right)^{2} \hat{P}^{(j) 2}}} \right] =&~ 0,
		\end{aligned}
	\end{align}
	allowing us to apply either charge before Gaussian apodization and propagation,
	\begin{align}
		\begin{aligned}
			\hat{\mathcal{Q}}^{(j)} \Phi_{\mathrm{G}}^{(j)} \left( z \right) =&~ e^{i z \sqrt{k^{2} - \left( k_{\perp}^{(j)} \right)^{2} \hat{P}^{(j) 2}}} \\
			&\times G^{(j)} \hat{\mathcal{Q}}^{(j)} \Phi^{(j)} \left( 0 \right), \\
			\hat{\mathcal{Q}}^{(j)\dagger} \Phi_{\mathrm{G}}^{(j)} \left( z \right) =&~ e^{i z \sqrt{k^{2} - \left( k_{\perp}^{(j)} \right)^{2} \hat{P}^{(j) 2}}} \\
			&\times G^{(j)} \hat{\mathcal{Q}}^{(j)\dagger} \Phi^{(j)} \left( 0 \right).
		\end{aligned}
	\end{align}
	
	For numerical propagation, we multiply the multishell seed and both non-Hermitian-charge partners by a Gaussian of amplitude radius $w_{0} = 3$ mm centered on their common rotation axis.
	We propagate the three complete inputs independently through their transverse Fourier spectra using the exact forward propagator,
	\begin{align}
		\begin{aligned}
			\Phi_{\chi,\mathrm{G}} \left( z \right) =&~ \mathcal{F}_{\perp}^{-1} \Big[ e^{i z \sqrt{k^{2} - k_{x}^{2} - k_{y}^{2}}} \\
			& \times \mathcal{F}_{\perp} \left[ G \Phi_{\chi} \left( 0 \right) \right] \Big],
		\end{aligned}
	\end{align}
	where the field label $\chi \in \{ \mathrm{multi}, \mathrm{o}, \mathrm{e} \}$ identifies the seed and its non-Hermitian-charge partners, $\mathcal{F}_{\perp}$ is the transverse Fourier transform, and $k_{x}$ and $k_{y}$ are the transverse wavevector components.
	\begingroup
	Figure \ref{fig:Fig4} shows our experimental setup for observing apodized Talbot fields.
	We expanded and collimated a low-power He--Ne laser beam to illuminate a phase--only spatial light modulator (SLM, Holoeye Pluto VIS).
	We encoded the Talbot fields into the first diffraction order of a digital hologram and isolated the desired order using a 4f system with a spatial filter placed at the back focal plane of L2.
	The 4f--system also imaged the SLM plane onto the back focal plane of L3, which we defined as the origin of propagation ($z=0$).
	Finally, we recorded the transverse intensity distributions at different propagation distances, $z / z_{\mathrm{T}} = 0, 1/8, 1/4, 1$, using a CCD camera (Zelux CS165MU1).
	
	\begin{figure}[t]
		\centering
		\includegraphics[width=86mm]{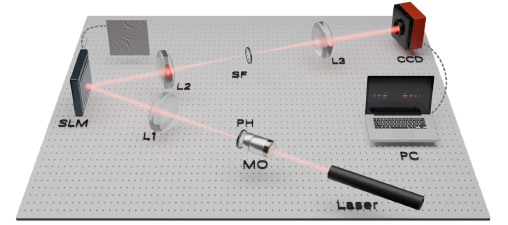}
		\caption{Experimental arrangement for generating and recording the Gaussian-apodized Talbot seed and its supersymmetric partners.
			The He--Ne laser illuminates the spatial light modulator (SLM) after expansion and collimation via the microscope objective (MO), pinhole (P), and lens L1. Lenses L2 and L3 relay the field to the $z = 0$ plane, with the spatial filter (SF) between them. The camera (CCD) records the intensity, and the computer (PC) displays the acquired image.}
		\label{fig:Fig4}
	\end{figure}
	
	Figure~\ref{fig:Fig5} shows the experimental intensities of the Gaussian-apodized seed and its two non-Hermitian-charge partners at $z / z_{\mathrm{T}} = 0, 1/8, 1/4, 1$.
	In the camera recordings at $z = 0$, the fields exhibit two dominant intensity regions along the horizontal axis, surrounded by weaker interference fringes.
	As the fields propagate to $z=z_T/8$, the dominant lobes broaden and redistribute their intensity, while the surrounding fringes become more pronounced.
	At $z=z_T/4$, the intensity distributions feature dominant lobes and interference fringes predominantly aligned along the vertical axis.
	The recorded intensities show the same qualitative symmetry as the numerical profiles.
	At $z=z_T$, the fields approximately recover their initial transverse intensity distributions, demonstrating the Talbot revival despite the finite apodization and experimental imperfections.
	Overall, the measured evolution agrees qualitatively with the numerical propagation predictions, reproducing the redistribution of intensity, the transformation of the interference fringes, and the revival of the initial patterns.
	\endgroup
	
	\begin{figure}[t]
		\centering
		\includegraphics[width=86mm]{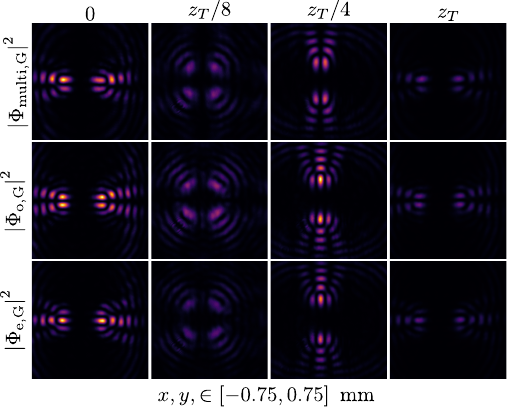}
		\caption{Experimental intensities of the Gaussian-apodized Talbot seed and its two non-Hermitian-charge partners for a Gaussian amplitude radius $w_{0} = 3$ mm.
			Rows show the intensities of $\Phi_{\mathrm{multi,G}}$, $\Phi_{\mathrm{o,G}}$, and $\Phi_{\mathrm{e,G}}$.
			Columns show $z / z_{\mathrm{T}} = 0, 1/8, 1/4, 1$.
			We normalize each row to its maximum intensity over the four planes and retain that scale throughout the row.
			Panels span $x, y \in [ -0.75, 0.75 ]$ mm, with the envelope centered at $(-0.1, 0)$ mm.}
		\label{fig:Fig5}
	\end{figure}
	
	%%%%%%%%%%%%%%%%%%%%%%%
	\section{Conclusions}
	\label{sec:Sec9}
	%%%%%%%%%%%%%%%%%%%%%%%
	The four separable families of the two-dimensional Helmholtz equation on a fixed transverse shell share reflection symmetry across the transverse horizontal axis, which defines a $\mathbb{Z}_{2}$ grading of the underlying Euclidean algebra.
	Within our graded algebra, we expressed all four separation operators through mixed supercommutators of displaced OAM generators, relating separation to the placement of rotation axes along the reflection axis.
	For Cartesian waves, we send the two rotation axes to opposite infinities; for Bessel waves, we place both at the origin; for Mathieu waves, we place them at the elliptic foci; and for Weber waves, we keep one at the origin and send the other to infinity, with the corresponding shifts and rescalings.
	
	We constructed supersymmetric charges connecting opposite-parity Cartesian and Bessel waves at equal separation eigenvalues by combining these generators with the reflection projectors.
	Our Cartesian charges shift the standing-wave pattern by a quarter period along the vertical axis, while our Bessel charges rotate the angular lobes by $\pi / (2\ell)$.
	For Mathieu waves at nonzero ellipticity, no odd operator preserves the separation eigenvalue, because the Mathieu spectrum is simple and its even and odd characteristic values do not coincide.
	For Weber waves, we constructed our charges from the product of the spectral sign of the vertical translation generator and the separation operator, exchanging reflection parity without changing the separation eigenvalue.
	Neither the Cartesian nor the Weber realization has a Fredholm Witten index because zero lies in the continuous spectrum of its super-Hamiltonian.
	The Bessel zero mode has no odd partner, giving Witten index one and unbroken supersymmetry for every rotation-axis position along the reflection line.
	
	On a single transverse shell, Helmholtz waves are propagation invariant.
	We extended our Bessel charges to multiple transverse shells to explore Talbot self-imaging of a seed and its supersymmetric partners.
	Through successive charge actions, we change the relative weights of the seed's OAM components while preserving the longitudinal phase differences, generating different spatial profiles with the same paraxial Talbot recurrence.
	We applied a Gaussian envelope centered on the common rotation axis to obtain finite-power fields.
	Our supersymmetric partner relations hold throughout free propagation, although the spectral broadening introduced by apodization modifies the ideal Talbot revival.
	We generated the Gaussian-apodized seed and its two partners with a phase-only spatial light modulator and recorded their transverse intensities during propagation.
	We observed the redistribution of the dominant lobes and interference fringes and the approximate recovery of the initial intensity profiles at the ideal Talbot distance.
	These observations agree qualitatively with our numerical propagation and support Talbot self-imaging of the finite-power fields.
	
	Our framework may guide the design and transformation of structured fields with controlled reflection parity and OAM content, with possible applications to mode conversion and optical encoding in distinct spatial profiles that share a prescribed self-imaging distance.
	
	%%%%%%%%%%%%%%%%%%%%%%% Backmatter %%%%%%%%%%%%%%%%%%%%%%%
	
	%%%%%%%%%%%%%%%%%%%%%%%
	\section*{Acknowledgements}
	%%%%%%%%%%%%%%%%%%%%%%%
	B.M.R.L. acknowledges the support and hospitality of the Department of Physics and Astronomy at the University of New Mexico as an affiliate visiting colleague.
	
	%%%%%%%%%%%%%%%%%%%%%%%
	\section*{Funding}
	%%%%%%%%%%%%%%%%%%%%%%%
	G.B. and B.P acknowledge support from the South African Quantum Technology Initiative (SAQuTI) and the Department of Science Technology and Innovation (DSTI).
	
	%%%%%%%%%%%%%%%%%%%%%%%
	\section*{Data Availability Statement}
	%%%%%%%%%%%%%%%%%%%%%%%
	The data that support the findings of this study are available from the corresponding author upon reasonable request.
	
	%%%%%%%%%%%%%%%%%%%%%%%
	\section*{Disclosures}
	%%%%%%%%%%%%%%%%%%%%%%%
	The authors declare no conflicts of interest.
	
	%%%%%%%%%%%%%%%%%%%%%%%
	\section*{Tool and Computational Resource}
	%%%%%%%%%%%%%%%%%%%%%%%
	Claude Science with the OpenAlex API was used for literature searches.
	OpenAI GPT-6 Astra and Claude Opus 5.5 were used as adversarial reviewers to improve the clarity of our exposition and our use of language.
	We take full responsibility for the accuracy, integrity, and attribution of the final work.
	
	%%%%%%%%%%%%%%%%%%%%%%% References %%%%%%%%%%%%%%%%%%%%%%%
	
%	\bibliographystyle{apsrev4-2}
%	\bibliography{references}

\begin{thebibliography}{49}%
	\makeatletter
	\providecommand \@ifxundefined [1]{%
		\@ifx{#1\undefined}
	}%
	\providecommand \@ifnum [1]{%
		\ifnum #1\expandafter \@firstoftwo
		\else \expandafter \@secondoftwo
		\fi
	}%
	\providecommand \@ifx [1]{%
		\ifx #1\expandafter \@firstoftwo
		\else \expandafter \@secondoftwo
		\fi
	}%
	\providecommand \natexlab [1]{#1}%
	\providecommand \enquote  [1]{``#1''}%
	\providecommand \bibnamefont  [1]{#1}%
	\providecommand \bibfnamefont [1]{#1}%
	\providecommand \citenamefont [1]{#1}%
	\providecommand \href@noop [0]{\@secondoftwo}%
	\providecommand \href [0]{\begingroup \@sanitize@url \@href}%
	\providecommand \@href[1]{\@@startlink{#1}\@@href}%
	\providecommand \@@href[1]{\endgroup#1\@@endlink}%
	\providecommand \@sanitize@url [0]{\catcode `\\12\catcode `\$12\catcode
		`\&12\catcode `\#12\catcode `\^12\catcode `\_12\catcode `\%12\relax}%
	\providecommand \@@startlink[1]{}%
	\providecommand \@@endlink[0]{}%
	\providecommand \url  [0]{\begingroup\@sanitize@url \@url }%
	\providecommand \@url [1]{\endgroup\@href {#1}{\urlprefix }}%
	\providecommand \urlprefix  [0]{URL }%
	\providecommand \Eprint [0]{\href }%
	\providecommand \doibase [0]{https://doi.org/}%
	\providecommand \selectlanguage [0]{\@gobble}%
	\providecommand \bibinfo  [0]{\@secondoftwo}%
	\providecommand \bibfield  [0]{\@secondoftwo}%
	\providecommand \translation [1]{[#1]}%
	\providecommand \BibitemOpen [0]{}%
	\providecommand \bibitemStop [0]{}%
	\providecommand \bibitemNoStop [0]{.\EOS\space}%
	\providecommand \EOS [0]{\spacefactor3000\relax}%
	\providecommand \BibitemShut  [1]{\csname bibitem#1\endcsname}%
	\let\auto@bib@innerbib\@empty
	%</preamble>
	\bibitem [{\citenamefont {Forbes}\ \emph {et~al.}(2021)\citenamefont {Forbes},
		\citenamefont {de~Oliveira},\ and\ \citenamefont {Dennis}}]{Forbes2021p253}%
	\BibitemOpen
	\bibfield  {author} {\bibinfo {author} {\bibfnamefont {A.}~\bibnamefont
			{Forbes}}, \bibinfo {author} {\bibfnamefont {M.}~\bibnamefont
			{de~Oliveira}},\ and\ \bibinfo {author} {\bibfnamefont {M.~R.}\ \bibnamefont
			{Dennis}},\ }\href {https://doi.org/10.1038/s41566-021-00780-4} {\bibfield
		{journal} {\bibinfo  {journal} {Nat. Photonics}\ }\textbf {\bibinfo {volume}
			{15}},\ \bibinfo {pages} {253} (\bibinfo {year} {2021})}\BibitemShut
	{NoStop}%
	\bibitem [{\citenamefont {Stratton}(1941)}]{Stratton1941}%
	\BibitemOpen
	\bibfield  {author} {\bibinfo {author} {\bibfnamefont {J.~A.}\ \bibnamefont
			{Stratton}},\ }\href@noop {} {\emph {\bibinfo {title} {Electromagnetic
				theory}}}\ (\bibinfo  {publisher} {McGraw-Hill},\ \bibinfo {address} {New
		York},\ \bibinfo {year} {1941})\ \bibinfo {note} {sec.~6.1,
		pp.~349--351}\BibitemShut {NoStop}%
	\bibitem [{\citenamefont {{Lord Rayleigh}}(1897)}]{Rayleigh1897p125}%
	\BibitemOpen
	\bibfield  {author} {\bibinfo {author} {\bibnamefont {{Lord Rayleigh}}},\
	}\href {https://doi.org/10.1080/14786449708620969} {\bibfield  {journal}
		{\bibinfo  {journal} {Philos. Mag. Ser.~5}\ }\textbf {\bibinfo {volume}
			{43}},\ \bibinfo {pages} {125} (\bibinfo {year} {1897})}\BibitemShut
	{NoStop}%
	\bibitem [{\citenamefont {Rodr{\'i}guez-Lara}\ and\ \citenamefont
		{J{\'a}uregui}(2008)}]{RodriguezLara2008p033813}%
	\BibitemOpen
	\bibfield  {author} {\bibinfo {author} {\bibfnamefont {B.~M.}\ \bibnamefont
			{Rodr{\'i}guez-Lara}}\ and\ \bibinfo {author} {\bibfnamefont
			{R.}~\bibnamefont {J{\'a}uregui}},\ }\href
	{https://doi.org/10.1103/PhysRevA.78.033813} {\bibfield  {journal} {\bibinfo
			{journal} {Phys. Rev. A}\ }\textbf {\bibinfo {volume} {78}},\ \bibinfo
		{pages} {033813} (\bibinfo {year} {2008})}\BibitemShut {NoStop}%
	\bibitem [{\citenamefont {Rodr{\'i}guez-Lara}\ and\ \citenamefont
		{J{\'a}uregui}(2009)}]{RodriguezLara2009055806}%
	\BibitemOpen
	\bibfield  {author} {\bibinfo {author} {\bibfnamefont {B.~M.}\ \bibnamefont
			{Rodr{\'i}guez-Lara}}\ and\ \bibinfo {author} {\bibfnamefont
			{R.}~\bibnamefont {J{\'a}uregui}},\ }\href
	{https://doi.org/10.1103/PhysRevA.79.055806} {\bibfield  {journal} {\bibinfo
			{journal} {Phys. Rev. A}\ }\textbf {\bibinfo {volume} {79}},\ \bibinfo
		{pages} {055806} (\bibinfo {year} {2009})}\BibitemShut {NoStop}%
	\bibitem [{\citenamefont {Durnin}(1987)}]{Durnin1987p651}%
	\BibitemOpen
	\bibfield  {author} {\bibinfo {author} {\bibfnamefont {J.}~\bibnamefont
			{Durnin}},\ }\href {https://doi.org/10.1364/JOSAA.4.000651} {\bibfield
		{journal} {\bibinfo  {journal} {J. Opt. Soc. Am. A}\ }\textbf {\bibinfo
			{volume} {4}},\ \bibinfo {pages} {651} (\bibinfo {year} {1987})}\BibitemShut
	{NoStop}%
	\bibitem [{\citenamefont {Miller}(1974)}]{Miller1974p626}%
	\BibitemOpen
	\bibfield  {author} {\bibinfo {author} {\bibfnamefont {W.}~\bibnamefont
			{Miller}, \bibfnamefont {Jr.}},\ }\href {https://doi.org/10.1137/0505063}
	{\bibfield  {journal} {\bibinfo  {journal} {SIAM J. Math. Anal.}\ }\textbf
		{\bibinfo {volume} {5}},\ \bibinfo {pages} {626} (\bibinfo {year}
		{1974})}\BibitemShut {NoStop}%
	\bibitem [{\citenamefont {Boyer}\ \emph {et~al.}(1976)\citenamefont {Boyer},
		\citenamefont {Kalnins},\ and\ \citenamefont {Miller}}]{Boyer1976p35}%
	\BibitemOpen
	\bibfield  {author} {\bibinfo {author} {\bibfnamefont {C.~P.}\ \bibnamefont
			{Boyer}}, \bibinfo {author} {\bibfnamefont {E.~G.}\ \bibnamefont {Kalnins}},\
		and\ \bibinfo {author} {\bibfnamefont {W.}~\bibnamefont {Miller},
			\bibfnamefont {Jr.}},\ }\href {https://doi.org/10.1017/S0027763000017165}
	{\bibfield  {journal} {\bibinfo  {journal} {Nagoya Math. J.}\ }\textbf
		{\bibinfo {volume} {60}},\ \bibinfo {pages} {35} (\bibinfo {year}
		{1976})}\BibitemShut {NoStop}%
	\bibitem [{\citenamefont {Miller}(1977)}]{Miller1977}%
	\BibitemOpen
	\bibfield  {author} {\bibinfo {author} {\bibfnamefont {W.}~\bibnamefont
			{Miller}, \bibfnamefont {Jr.}},\ }\href@noop {} {\emph {\bibinfo {title}
			{Symmetry and separation of variables}}},\ \bibinfo {series} {Encyclopedia of
		Mathematics and its Applications}, Vol.~\bibinfo {volume} {4}\ (\bibinfo
	{publisher} {Addison-Wesley},\ \bibinfo {address} {Reading, Massachusetts},\
	\bibinfo {year} {1977})\BibitemShut {NoStop}%
	\bibitem [{\citenamefont {Whittaker}(1903)}]{Whittaker1903p333}%
	\BibitemOpen
	\bibfield  {author} {\bibinfo {author} {\bibfnamefont {E.~T.}\ \bibnamefont
			{Whittaker}},\ }\href {https://doi.org/10.1007/BF01444290} {\bibfield
		{journal} {\bibinfo  {journal} {Math. Ann.}\ }\textbf {\bibinfo {volume}
			{57}},\ \bibinfo {pages} {333} (\bibinfo {year} {1903})}\BibitemShut
	{NoStop}%
	\bibitem [{\citenamefont {Durnin}\ \emph {et~al.}(1987)\citenamefont {Durnin},
		\citenamefont {Miceli},\ and\ \citenamefont {Eberly}}]{Durnin1987p1499}%
	\BibitemOpen
	\bibfield  {author} {\bibinfo {author} {\bibfnamefont {J.}~\bibnamefont
			{Durnin}}, \bibinfo {author} {\bibfnamefont {J.~J.}\ \bibnamefont {Miceli},
			\bibfnamefont {Jr.}},\ and\ \bibinfo {author} {\bibfnamefont {J.~H.}\
			\bibnamefont {Eberly}},\ }\href {https://doi.org/10.1103/PhysRevLett.58.1499}
	{\bibfield  {journal} {\bibinfo  {journal} {Phys. Rev. Lett.}\ }\textbf
		{\bibinfo {volume} {58}},\ \bibinfo {pages} {1499} (\bibinfo {year}
		{1987})}\BibitemShut {NoStop}%
	\bibitem [{\citenamefont {Garc{\'e}s-Ch{\'a}vez}\ \emph
		{et~al.}(2002)\citenamefont {Garc{\'e}s-Ch{\'a}vez}, \citenamefont {McGloin},
		\citenamefont {Melville}, \citenamefont {Sibbett},\ and\ \citenamefont
		{Dholakia}}]{GarcesChavez2002p145}%
	\BibitemOpen
	\bibfield  {author} {\bibinfo {author} {\bibfnamefont {V.}~\bibnamefont
			{Garc{\'e}s-Ch{\'a}vez}}, \bibinfo {author} {\bibfnamefont {D.}~\bibnamefont
			{McGloin}}, \bibinfo {author} {\bibfnamefont {H.}~\bibnamefont {Melville}},
		\bibinfo {author} {\bibfnamefont {W.}~\bibnamefont {Sibbett}},\ and\ \bibinfo
		{author} {\bibfnamefont {K.}~\bibnamefont {Dholakia}},\ }\href
	{https://doi.org/10.1038/nature01007} {\bibfield  {journal} {\bibinfo
			{journal} {Nature}\ }\textbf {\bibinfo {volume} {419}},\ \bibinfo {pages}
		{145} (\bibinfo {year} {2002})}\BibitemShut {NoStop}%
	\bibitem [{\citenamefont {Fahrbach}\ \emph {et~al.}(2010)\citenamefont
		{Fahrbach}, \citenamefont {Simon},\ and\ \citenamefont
		{Rohrbach}}]{Fahrbach2010p780}%
	\BibitemOpen
	\bibfield  {author} {\bibinfo {author} {\bibfnamefont {F.~O.}\ \bibnamefont
			{Fahrbach}}, \bibinfo {author} {\bibfnamefont {P.}~\bibnamefont {Simon}},\
		and\ \bibinfo {author} {\bibfnamefont {A.}~\bibnamefont {Rohrbach}},\ }\href
	{https://doi.org/10.1038/nphoton.2010.204} {\bibfield  {journal} {\bibinfo
			{journal} {Nat. Photonics}\ }\textbf {\bibinfo {volume} {4}},\ \bibinfo
		{pages} {780} (\bibinfo {year} {2010})}\BibitemShut {NoStop}%
	\bibitem [{\citenamefont {Planchon}\ \emph {et~al.}(2011)\citenamefont
		{Planchon}, \citenamefont {Gao}, \citenamefont {Milkie}, \citenamefont
		{Davidson}, \citenamefont {Galbraith}, \citenamefont {Galbraith},\ and\
		\citenamefont {Betzig}}]{Planchon2011p417}%
	\BibitemOpen
	\bibfield  {author} {\bibinfo {author} {\bibfnamefont {T.~A.}\ \bibnamefont
			{Planchon}}, \bibinfo {author} {\bibfnamefont {L.}~\bibnamefont {Gao}},
		\bibinfo {author} {\bibfnamefont {D.~E.}\ \bibnamefont {Milkie}}, \bibinfo
		{author} {\bibfnamefont {M.~W.}\ \bibnamefont {Davidson}}, \bibinfo {author}
		{\bibfnamefont {J.~A.}\ \bibnamefont {Galbraith}}, \bibinfo {author}
		{\bibfnamefont {C.~G.}\ \bibnamefont {Galbraith}},\ and\ \bibinfo {author}
		{\bibfnamefont {E.}~\bibnamefont {Betzig}},\ }\href
	{https://doi.org/10.1038/nmeth.1586} {\bibfield  {journal} {\bibinfo
			{journal} {Nat. Methods}\ }\textbf {\bibinfo {volume} {8}},\ \bibinfo {pages}
		{417} (\bibinfo {year} {2011})}\BibitemShut {NoStop}%
	\bibitem [{\citenamefont {Volke-Sepulveda}\ \emph {et~al.}(2002)\citenamefont
		{Volke-Sepulveda}, \citenamefont {Garc{\'e}s-Ch{\'a}vez}, \citenamefont
		{Ch{\'a}vez-Cerda}, \citenamefont {Arlt},\ and\ \citenamefont
		{Dholakia}}]{VolkeSepulveda2002pS82}%
	\BibitemOpen
	\bibfield  {author} {\bibinfo {author} {\bibfnamefont {K.}~\bibnamefont
			{Volke-Sepulveda}}, \bibinfo {author} {\bibfnamefont {V.}~\bibnamefont
			{Garc{\'e}s-Ch{\'a}vez}}, \bibinfo {author} {\bibfnamefont {S.}~\bibnamefont
			{Ch{\'a}vez-Cerda}}, \bibinfo {author} {\bibfnamefont {J.}~\bibnamefont
			{Arlt}},\ and\ \bibinfo {author} {\bibfnamefont {K.}~\bibnamefont
			{Dholakia}},\ }\href {https://doi.org/10.1088/1464-4266/4/2/373} {\bibfield
		{journal} {\bibinfo  {journal} {J. Opt. B: Quantum Semiclass. Opt.}\ }\textbf
		{\bibinfo {volume} {4}},\ \bibinfo {pages} {S82} (\bibinfo {year}
		{2002})}\BibitemShut {NoStop}%
	\bibitem [{\citenamefont {J{\'a}uregui}\ and\ \citenamefont
		{Hacyan}(2005)}]{Jauregui2005p033411}%
	\BibitemOpen
	\bibfield  {author} {\bibinfo {author} {\bibfnamefont {R.}~\bibnamefont
			{J{\'a}uregui}}\ and\ \bibinfo {author} {\bibfnamefont {S.}~\bibnamefont
			{Hacyan}},\ }\href {https://doi.org/10.1103/PhysRevA.71.033411} {\bibfield
		{journal} {\bibinfo  {journal} {Phys. Rev. A}\ }\textbf {\bibinfo {volume}
			{71}},\ \bibinfo {pages} {033411} (\bibinfo {year} {2005})}\BibitemShut
	{NoStop}%
	\bibitem [{\citenamefont {Saltiel}\ \emph {et~al.}(2007)\citenamefont
		{Saltiel}, \citenamefont {Krolikowski}, \citenamefont {Neshev},\ and\
		\citenamefont {Kivshar}}]{Saltiel2007p4132}%
	\BibitemOpen
	\bibfield  {author} {\bibinfo {author} {\bibfnamefont {S.}~\bibnamefont
			{Saltiel}}, \bibinfo {author} {\bibfnamefont {W.}~\bibnamefont
			{Krolikowski}}, \bibinfo {author} {\bibfnamefont {D.}~\bibnamefont
			{Neshev}},\ and\ \bibinfo {author} {\bibfnamefont {Y.~S.}\ \bibnamefont
			{Kivshar}},\ }\href {https://doi.org/10.1364/OE.15.004132} {\bibfield
		{journal} {\bibinfo  {journal} {Opt. Express}\ }\textbf {\bibinfo {volume}
			{15}},\ \bibinfo {pages} {4132} (\bibinfo {year} {2007})}\BibitemShut
	{NoStop}%
	\bibitem [{\citenamefont {Trichili}\ \emph {et~al.}(2014)\citenamefont
		{Trichili}, \citenamefont {Mhlanga}, \citenamefont {Ismail}, \citenamefont
		{Roux}, \citenamefont {McLaren}, \citenamefont {Zghal},\ and\ \citenamefont
		{Forbes}}]{Trichili2014p17553}%
	\BibitemOpen
	\bibfield  {author} {\bibinfo {author} {\bibfnamefont {A.}~\bibnamefont
			{Trichili}}, \bibinfo {author} {\bibfnamefont {T.}~\bibnamefont {Mhlanga}},
		\bibinfo {author} {\bibfnamefont {Y.}~\bibnamefont {Ismail}}, \bibinfo
		{author} {\bibfnamefont {F.~S.}\ \bibnamefont {Roux}}, \bibinfo {author}
		{\bibfnamefont {M.}~\bibnamefont {McLaren}}, \bibinfo {author} {\bibfnamefont
			{M.}~\bibnamefont {Zghal}},\ and\ \bibinfo {author} {\bibfnamefont
			{A.}~\bibnamefont {Forbes}},\ }\href {https://doi.org/10.1364/OE.22.017553}
	{\bibfield  {journal} {\bibinfo  {journal} {Opt. Express}\ }\textbf {\bibinfo
			{volume} {22}},\ \bibinfo {pages} {17553} (\bibinfo {year}
		{2014})}\BibitemShut {NoStop}%
	\bibitem [{\citenamefont {Guti{\'e}rrez-Vega}\ \emph
		{et~al.}(2000)\citenamefont {Guti{\'e}rrez-Vega}, \citenamefont
		{Iturbe-Castillo},\ and\ \citenamefont
		{Ch{\'a}vez-Cerda}}]{GutierrezVega2000p1493}%
	\BibitemOpen
	\bibfield  {author} {\bibinfo {author} {\bibfnamefont {J.~C.}\ \bibnamefont
			{Guti{\'e}rrez-Vega}}, \bibinfo {author} {\bibfnamefont {M.~D.}\ \bibnamefont
			{Iturbe-Castillo}},\ and\ \bibinfo {author} {\bibfnamefont {S.}~\bibnamefont
			{Ch{\'a}vez-Cerda}},\ }\href {https://doi.org/10.1364/OL.25.001493}
	{\bibfield  {journal} {\bibinfo  {journal} {Opt. Lett.}\ }\textbf {\bibinfo
			{volume} {25}},\ \bibinfo {pages} {1493} (\bibinfo {year}
		{2000})}\BibitemShut {NoStop}%
	\bibitem [{\citenamefont {Ch{\'a}vez-Cerda}\ \emph {et~al.}(2001)\citenamefont
		{Ch{\'a}vez-Cerda}, \citenamefont {Guti{\'e}rrez-Vega},\ and\ \citenamefont
		{New}}]{ChavezCerda2001p1803}%
	\BibitemOpen
	\bibfield  {author} {\bibinfo {author} {\bibfnamefont {S.}~\bibnamefont
			{Ch{\'a}vez-Cerda}}, \bibinfo {author} {\bibfnamefont {J.~C.}\ \bibnamefont
			{Guti{\'e}rrez-Vega}},\ and\ \bibinfo {author} {\bibfnamefont {G.~H.~C.}\
			\bibnamefont {New}},\ }\href {https://doi.org/10.1364/OL.26.001803}
	{\bibfield  {journal} {\bibinfo  {journal} {Opt. Lett.}\ }\textbf {\bibinfo
			{volume} {26}},\ \bibinfo {pages} {1803} (\bibinfo {year}
		{2001})}\BibitemShut {NoStop}%
	\bibitem [{\citenamefont {Chaos-Cador}\ and\ \citenamefont
		{Ley-Koo}(2002)}]{ChaosCador2002p67}%
	\BibitemOpen
	\bibfield  {author} {\bibinfo {author} {\bibfnamefont {L.}~\bibnamefont
			{Chaos-Cador}}\ and\ \bibinfo {author} {\bibfnamefont {E.}~\bibnamefont
			{Ley-Koo}},\ }\href {https://rmf.smf.mx/ojs/index.php/rmf/article/view/3035}
	{\bibfield  {journal} {\bibinfo  {journal} {Rev. Mex. F\'is.}\ }\textbf
		{\bibinfo {volume} {48}},\ \bibinfo {pages} {67} (\bibinfo {year}
		{2002})}\BibitemShut {NoStop}%
	\bibitem [{\citenamefont {Ch{\'a}vez-Cerda}\ \emph {et~al.}(2002)\citenamefont
		{Ch{\'a}vez-Cerda}, \citenamefont {Padgett}, \citenamefont {Allison},
		\citenamefont {New}, \citenamefont {Guti{\'e}rrez-Vega}, \citenamefont
		{O'Neil}, \citenamefont {MacVicar},\ and\ \citenamefont
		{Courtial}}]{ChavezCerda2002pS52}%
	\BibitemOpen
	\bibfield  {author} {\bibinfo {author} {\bibfnamefont {S.}~\bibnamefont
			{Ch{\'a}vez-Cerda}}, \bibinfo {author} {\bibfnamefont {M.~J.}\ \bibnamefont
			{Padgett}}, \bibinfo {author} {\bibfnamefont {I.}~\bibnamefont {Allison}},
		\bibinfo {author} {\bibfnamefont {G.~H.~C.}\ \bibnamefont {New}}, \bibinfo
		{author} {\bibfnamefont {J.~C.}\ \bibnamefont {Guti{\'e}rrez-Vega}}, \bibinfo
		{author} {\bibfnamefont {A.~T.}\ \bibnamefont {O'Neil}}, \bibinfo {author}
		{\bibfnamefont {I.}~\bibnamefont {MacVicar}},\ and\ \bibinfo {author}
		{\bibfnamefont {J.}~\bibnamefont {Courtial}},\ }\href
	{https://doi.org/10.1088/1464-4266/4/2/368} {\bibfield  {journal} {\bibinfo
			{journal} {J. Opt. B: Quantum Semiclass. Opt.}\ }\textbf {\bibinfo {volume}
			{4}},\ \bibinfo {pages} {S52} (\bibinfo {year} {2002})}\BibitemShut {NoStop}%
	\bibitem [{\citenamefont {Bandres}\ \emph {et~al.}(2004)\citenamefont
		{Bandres}, \citenamefont {Guti{\'e}rrez-Vega},\ and\ \citenamefont
		{Ch{\'a}vez-Cerda}}]{Bandres2004p44}%
	\BibitemOpen
	\bibfield  {author} {\bibinfo {author} {\bibfnamefont {M.~A.}\ \bibnamefont
			{Bandres}}, \bibinfo {author} {\bibfnamefont {J.~C.}\ \bibnamefont
			{Guti{\'e}rrez-Vega}},\ and\ \bibinfo {author} {\bibfnamefont
			{S.}~\bibnamefont {Ch{\'a}vez-Cerda}},\ }\href
	{https://doi.org/10.1364/OL.29.000044} {\bibfield  {journal} {\bibinfo
			{journal} {Opt. Lett.}\ }\textbf {\bibinfo {volume} {29}},\ \bibinfo {pages}
		{44} (\bibinfo {year} {2004})}\BibitemShut {NoStop}%
	\bibitem [{\citenamefont {L{\'o}pez-Mariscal}\ \emph
		{et~al.}(2005)\citenamefont {L{\'o}pez-Mariscal}, \citenamefont {Bandres},
		\citenamefont {Guti{\'e}rrez-Vega},\ and\ \citenamefont
		{Ch{\'a}vez-Cerda}}]{LopezMariscal2005p2364}%
	\BibitemOpen
	\bibfield  {author} {\bibinfo {author} {\bibfnamefont {C.}~\bibnamefont
			{L{\'o}pez-Mariscal}}, \bibinfo {author} {\bibfnamefont {M.~A.}\ \bibnamefont
			{Bandres}}, \bibinfo {author} {\bibfnamefont {J.~C.}\ \bibnamefont
			{Guti{\'e}rrez-Vega}},\ and\ \bibinfo {author} {\bibfnamefont
			{S.}~\bibnamefont {Ch{\'a}vez-Cerda}},\ }\href
	{https://doi.org/10.1364/OPEX.13.002364} {\bibfield  {journal} {\bibinfo
			{journal} {Opt. Express}\ }\textbf {\bibinfo {volume} {13}},\ \bibinfo
		{pages} {2364} (\bibinfo {year} {2005})}\BibitemShut {NoStop}%
	\bibitem [{\citenamefont {Hern{\'a}ndez-Cedillo}\ \emph
		{et~al.}(2013)\citenamefont {Hern{\'a}ndez-Cedillo}, \citenamefont {Bernon},
		\citenamefont {Hattermann}, \citenamefont {Fort{\'a}gh},\ and\ \citenamefont
		{J{\'a}uregui}}]{HernandezCedillo2013p023404}%
	\BibitemOpen
	\bibfield  {author} {\bibinfo {author} {\bibfnamefont {C.~L.}\ \bibnamefont
			{Hern{\'a}ndez-Cedillo}}, \bibinfo {author} {\bibfnamefont {S.}~\bibnamefont
			{Bernon}}, \bibinfo {author} {\bibfnamefont {H.}~\bibnamefont {Hattermann}},
		\bibinfo {author} {\bibfnamefont {J.}~\bibnamefont {Fort{\'a}gh}},\ and\
		\bibinfo {author} {\bibfnamefont {R.}~\bibnamefont {J{\'a}uregui}},\ }\href
	{https://doi.org/10.1103/PhysRevA.87.023404} {\bibfield  {journal} {\bibinfo
			{journal} {Phys. Rev. A}\ }\textbf {\bibinfo {volume} {87}},\ \bibinfo
		{pages} {023404} (\bibinfo {year} {2013})}\BibitemShut {NoStop}%
	\bibitem [{\citenamefont {Bandres}\ and\ \citenamefont
		{Rodr{\'i}guez-Lara}(2013)}]{Bandres2013p013054}%
	\BibitemOpen
	\bibfield  {author} {\bibinfo {author} {\bibfnamefont {M.~A.}\ \bibnamefont
			{Bandres}}\ and\ \bibinfo {author} {\bibfnamefont {B.~M.}\ \bibnamefont
			{Rodr{\'i}guez-Lara}},\ }\href
	{https://doi.org/10.1088/1367-2630/15/1/013054} {\bibfield  {journal}
		{\bibinfo  {journal} {New J. Phys.}\ }\textbf {\bibinfo {volume} {15}},\
		\bibinfo {pages} {013054} (\bibinfo {year} {2013})}\BibitemShut {NoStop}%
	\bibitem [{\citenamefont {Rodr{\'i}guez-Lara}(2010)}]{RodriguezLara2010p327}%
	\BibitemOpen
	\bibfield  {author} {\bibinfo {author} {\bibfnamefont {B.~M.}\ \bibnamefont
			{Rodr{\'i}guez-Lara}},\ }\href {https://doi.org/10.1364/JOSAA.27.000327}
	{\bibfield  {journal} {\bibinfo  {journal} {J. Opt. Soc. Am. A}\ }\textbf
		{\bibinfo {volume} {27}},\ \bibinfo {pages} {327} (\bibinfo {year}
		{2010})}\BibitemShut {NoStop}%
	\bibitem [{\citenamefont {Dom{\'i}nguez-Cruz}\ and\ \citenamefont
		{Guti{\'e}rrez-Vega}(2026{\natexlab{a}})}]{DominguezCruz2026p055601}%
	\BibitemOpen
	\bibfield  {author} {\bibinfo {author} {\bibfnamefont {A.}~\bibnamefont
			{Dom{\'i}nguez-Cruz}}\ and\ \bibinfo {author} {\bibfnamefont {J.~C.}\
			\bibnamefont {Guti{\'e}rrez-Vega}},\ }\href
	{https://doi.org/10.1088/2040-8986/ae6460} {\bibfield  {journal} {\bibinfo
			{journal} {J. Opt.}\ }\textbf {\bibinfo {volume} {28}},\ \bibinfo {pages}
		{055601} (\bibinfo {year} {2026}{\natexlab{a}})}\BibitemShut {NoStop}%
	\bibitem [{\citenamefont {Dom{\'i}nguez-Cruz}\ and\ \citenamefont
		{Guti{\'e}rrez-Vega}(2026{\natexlab{b}})}]{DominguezCruz2026ae929a}%
	\BibitemOpen
	\bibfield  {author} {\bibinfo {author} {\bibfnamefont {A.}~\bibnamefont
			{Dom{\'i}nguez-Cruz}}\ and\ \bibinfo {author} {\bibfnamefont {J.~C.}\
			\bibnamefont {Guti{\'e}rrez-Vega}},\ }\href
	{https://doi.org/10.1088/2040-8986/ae929a} {\bibfield  {journal} {\bibinfo
			{journal} {J. Opt.}\ }\textbf {\bibinfo {volume} {28}},\ \bibinfo {pages}
		{085602} (\bibinfo {year} {2026}{\natexlab{b}})}\BibitemShut {NoStop}%
	\bibitem [{\citenamefont {Snyder}\ and\ \citenamefont
		{Love}(1983)}]{Snyder1983}%
	\BibitemOpen
	\bibfield  {author} {\bibinfo {author} {\bibfnamefont {A.~W.}\ \bibnamefont
			{Snyder}}\ and\ \bibinfo {author} {\bibfnamefont {J.~D.}\ \bibnamefont
			{Love}},\ }\href {https://doi.org/10.1007/978-1-4613-2813-1} {\emph {\bibinfo
			{title} {Optical waveguide theory}}}\ (\bibinfo  {publisher} {Chapman and
		Hall},\ \bibinfo {address} {London},\ \bibinfo {year} {1983})\BibitemShut
	{NoStop}%
	\bibitem [{\citenamefont {Chumakov}\ and\ \citenamefont
		{Wolf}(1994)}]{Chumakov1994p51}%
	\BibitemOpen
	\bibfield  {author} {\bibinfo {author} {\bibfnamefont {S.~M.}\ \bibnamefont
			{Chumakov}}\ and\ \bibinfo {author} {\bibfnamefont {K.~B.}\ \bibnamefont
			{Wolf}},\ }\href {https://doi.org/10.1016/0375-9601(94)00616-4} {\bibfield
		{journal} {\bibinfo  {journal} {Phys. Lett. A}\ }\textbf {\bibinfo {volume}
			{193}},\ \bibinfo {pages} {51} (\bibinfo {year} {1994})}\BibitemShut
	{NoStop}%
	\bibitem [{\citenamefont {Miri}\ \emph {et~al.}(2013)\citenamefont {Miri},
		\citenamefont {Heinrich}, \citenamefont {El-Ganainy},\ and\ \citenamefont
		{Christodoulides}}]{Miri2013p233902}%
	\BibitemOpen
	\bibfield  {author} {\bibinfo {author} {\bibfnamefont {M.-A.}\ \bibnamefont
			{Miri}}, \bibinfo {author} {\bibfnamefont {M.}~\bibnamefont {Heinrich}},
		\bibinfo {author} {\bibfnamefont {R.}~\bibnamefont {El-Ganainy}},\ and\
		\bibinfo {author} {\bibfnamefont {D.~N.}\ \bibnamefont {Christodoulides}},\
	}\href {https://doi.org/10.1103/PhysRevLett.110.233902} {\bibfield  {journal}
		{\bibinfo  {journal} {Phys. Rev. Lett.}\ }\textbf {\bibinfo {volume} {110}},\
		\bibinfo {pages} {233902} (\bibinfo {year} {2013})}\BibitemShut {NoStop}%
	\bibitem [{\citenamefont {Heinrich}\ \emph {et~al.}(2014)\citenamefont
		{Heinrich}, \citenamefont {Miri}, \citenamefont {St{\"u}tzer}, \citenamefont
		{El-Ganainy}, \citenamefont {Nolte}, \citenamefont {Szameit},\ and\
		\citenamefont {Christodoulides}}]{Heinrich2014p3698}%
	\BibitemOpen
	\bibfield  {author} {\bibinfo {author} {\bibfnamefont {M.}~\bibnamefont
			{Heinrich}}, \bibinfo {author} {\bibfnamefont {M.-A.}\ \bibnamefont {Miri}},
		\bibinfo {author} {\bibfnamefont {S.}~\bibnamefont {St{\"u}tzer}}, \bibinfo
		{author} {\bibfnamefont {R.}~\bibnamefont {El-Ganainy}}, \bibinfo {author}
		{\bibfnamefont {S.}~\bibnamefont {Nolte}}, \bibinfo {author} {\bibfnamefont
			{A.}~\bibnamefont {Szameit}},\ and\ \bibinfo {author} {\bibfnamefont {D.~N.}\
			\bibnamefont {Christodoulides}},\ }\href {https://doi.org/10.1038/ncomms4698}
	{\bibfield  {journal} {\bibinfo  {journal} {Nat. Commun.}\ }\textbf {\bibinfo
			{volume} {5}},\ \bibinfo {pages} {3698} (\bibinfo {year} {2014})}\BibitemShut
	{NoStop}%
	\bibitem [{\citenamefont {Zu{\~n}iga-Segundo}\ \emph
		{et~al.}(2014)\citenamefont {Zu{\~n}iga-Segundo}, \citenamefont
		{Rodr{\'\i}guez-Lara}, \citenamefont {{Fern\'andez C.}},\ and\ \citenamefont
		{Moya-Cessa}}]{ZunigaSegundo2014p987}%
	\BibitemOpen
	\bibfield  {author} {\bibinfo {author} {\bibfnamefont {A.}~\bibnamefont
			{Zu{\~n}iga-Segundo}}, \bibinfo {author} {\bibfnamefont {B.~M.}\ \bibnamefont
			{Rodr{\'\i}guez-Lara}}, \bibinfo {author} {\bibfnamefont {D.~J.}\
			\bibnamefont {{Fern\'andez C.}}},\ and\ \bibinfo {author} {\bibfnamefont
			{H.~M.}\ \bibnamefont {Moya-Cessa}},\ }\href
	{https://doi.org/10.1364/oe.22.000987} {\bibfield  {journal} {\bibinfo
			{journal} {Opt. Express}\ }\textbf {\bibinfo {volume} {22}},\ \bibinfo
		{pages} {987} (\bibinfo {year} {2014})}\BibitemShut {NoStop}%
	\bibitem [{\citenamefont {{Martinez Berumen}}\ and\ \citenamefont
		{Rodr{\'\i}guez-Lara}(2020)}]{MartinezBerumen2020p043521}%
	\BibitemOpen
	\bibfield  {author} {\bibinfo {author} {\bibfnamefont {P.~I.}\ \bibnamefont
			{{Martinez Berumen}}}\ and\ \bibinfo {author} {\bibfnamefont {B.~M.}\
			\bibnamefont {Rodr{\'\i}guez-Lara}},\ }\href
	{https://doi.org/10.1103/physreva.102.043521} {\bibfield  {journal} {\bibinfo
			{journal} {Phys. Rev. A}\ }\textbf {\bibinfo {volume} {102}},\ \bibinfo
		{pages} {043521} (\bibinfo {year} {2020})}\BibitemShut {NoStop}%
	\bibitem [{\citenamefont {Miri}\ \emph {et~al.}(2014)\citenamefont {Miri},
		\citenamefont {Heinrich},\ and\ \citenamefont
		{Christodoulides}}]{Miri2014p89}%
	\BibitemOpen
	\bibfield  {author} {\bibinfo {author} {\bibfnamefont {M.-A.}\ \bibnamefont
			{Miri}}, \bibinfo {author} {\bibfnamefont {M.}~\bibnamefont {Heinrich}},\
		and\ \bibinfo {author} {\bibfnamefont {D.~N.}\ \bibnamefont
			{Christodoulides}},\ }\href {https://doi.org/10.1364/OPTICA.1.000089}
	{\bibfield  {journal} {\bibinfo  {journal} {Optica}\ }\textbf {\bibinfo
			{volume} {1}},\ \bibinfo {pages} {89} (\bibinfo {year} {2014})}\BibitemShut
	{NoStop}%
	\bibitem [{\citenamefont {Hokmabadi}\ \emph {et~al.}(2019)\citenamefont
		{Hokmabadi}, \citenamefont {Nye}, \citenamefont {El-Ganainy}, \citenamefont
		{Christodoulides},\ and\ \citenamefont {Khajavikhan}}]{Hokmabadi2019p623}%
	\BibitemOpen
	\bibfield  {author} {\bibinfo {author} {\bibfnamefont {M.~P.}\ \bibnamefont
			{Hokmabadi}}, \bibinfo {author} {\bibfnamefont {N.~S.}\ \bibnamefont {Nye}},
		\bibinfo {author} {\bibfnamefont {R.}~\bibnamefont {El-Ganainy}}, \bibinfo
		{author} {\bibfnamefont {D.~N.}\ \bibnamefont {Christodoulides}},\ and\
		\bibinfo {author} {\bibfnamefont {M.}~\bibnamefont {Khajavikhan}},\ }\href
	{https://doi.org/10.1126/science.aav5103} {\bibfield  {journal} {\bibinfo
			{journal} {Science}\ }\textbf {\bibinfo {volume} {363}},\ \bibinfo {pages}
		{623} (\bibinfo {year} {2019})}\BibitemShut {NoStop}%
	\bibitem [{\citenamefont {Chandra}\ and\ \citenamefont
		{Litchinitser}(2021)}]{Chandra2021p59}%
	\BibitemOpen
	\bibfield  {author} {\bibinfo {author} {\bibfnamefont {N.}~\bibnamefont
			{Chandra}}\ and\ \bibinfo {author} {\bibfnamefont {N.~M.}\ \bibnamefont
			{Litchinitser}},\ }\href {https://doi.org/10.1038/s42005-021-00560-0}
	{\bibfield  {journal} {\bibinfo  {journal} {Commun. Phys.}\ }\textbf
		{\bibinfo {volume} {4}},\ \bibinfo {pages} {59} (\bibinfo {year}
		{2021})}\BibitemShut {NoStop}%
	\bibitem [{\citenamefont {Dom{\'i}nguez-Cruz}\ and\ \citenamefont
		{Rodr{\'\i}guez-Lara}(2026)}]{DominguezCruz2026p2606.26068}%
	\BibitemOpen
	\bibfield  {author} {\bibinfo {author} {\bibfnamefont {A.}~\bibnamefont
			{Dom{\'i}nguez-Cruz}}\ and\ \bibinfo {author} {\bibfnamefont {B.~M.}\
			\bibnamefont {Rodr{\'\i}guez-Lara}},\ }\href
	{https://doi.org/10.1103/hflr-6ksj} {\bibfield  {journal} {\bibinfo
			{journal} {Phys. Rev. A}\ }\textbf {\bibinfo {volume} {114}},\ \bibinfo
		{pages} {043526} (\bibinfo {year} {2026})}\BibitemShut {NoStop}%
	\bibitem [{\citenamefont {Kac}(1977)}]{Kac1977p8}%
	\BibitemOpen
	\bibfield  {author} {\bibinfo {author} {\bibfnamefont {V.~G.}\ \bibnamefont
			{Kac}},\ }\href {https://doi.org/10.1016/0001-8708(77)90017-2} {\bibfield
		{journal} {\bibinfo  {journal} {Adv. Math.}\ }\textbf {\bibinfo {volume}
			{26}},\ \bibinfo {pages} {8} (\bibinfo {year} {1977})}\BibitemShut {NoStop}%
	\bibitem [{\citenamefont {Erikson}\ and\ \citenamefont
		{Hill}(1949)}]{Erikson1949p29}%
	\BibitemOpen
	\bibfield  {author} {\bibinfo {author} {\bibfnamefont {H.~A.}\ \bibnamefont
			{Erikson}}\ and\ \bibinfo {author} {\bibfnamefont {E.~L.}\ \bibnamefont
			{Hill}},\ }\href {https://doi.org/10.1103/PhysRev.75.29} {\bibfield
		{journal} {\bibinfo  {journal} {Phys. Rev.}\ }\textbf {\bibinfo {volume}
			{75}},\ \bibinfo {pages} {29} (\bibinfo {year} {1949})}\BibitemShut {NoStop}%
	\bibitem [{\citenamefont {Coulson}\ and\ \citenamefont
		{Joseph}(1967)}]{Coulson1967p337}%
	\BibitemOpen
	\bibfield  {author} {\bibinfo {author} {\bibfnamefont {C.~A.}\ \bibnamefont
			{Coulson}}\ and\ \bibinfo {author} {\bibfnamefont {A.}~\bibnamefont
			{Joseph}},\ }\href {https://doi.org/10.1002/qua.560010405} {\bibfield
		{journal} {\bibinfo  {journal} {Int. J. Quantum Chem.}\ }\textbf {\bibinfo
			{volume} {1}},\ \bibinfo {pages} {337} (\bibinfo {year} {1967})}\BibitemShut
	{NoStop}%
	\bibitem [{\citenamefont {Izmest'ev}\ \emph {et~al.}(1996)\citenamefont
		{Izmest'ev}, \citenamefont {Pogosyan}, \citenamefont {Sissakian},\ and\
		\citenamefont {Winternitz}}]{Izmestev1996p5949}%
	\BibitemOpen
	\bibfield  {author} {\bibinfo {author} {\bibfnamefont {A.~A.}\ \bibnamefont
			{Izmest'ev}}, \bibinfo {author} {\bibfnamefont {G.~S.}\ \bibnamefont
			{Pogosyan}}, \bibinfo {author} {\bibfnamefont {A.~N.}\ \bibnamefont
			{Sissakian}},\ and\ \bibinfo {author} {\bibfnamefont {P.}~\bibnamefont
			{Winternitz}},\ }\href {https://doi.org/10.1088/0305-4470/29/18/024}
	{\bibfield  {journal} {\bibinfo  {journal} {J. Phys. A: Math. Gen.}\ }\textbf
		{\bibinfo {volume} {29}},\ \bibinfo {pages} {5949} (\bibinfo {year}
		{1996})}\BibitemShut {NoStop}%
	\bibitem [{\citenamefont {Witten}(1982)}]{Witten1982p253}%
	\BibitemOpen
	\bibfield  {author} {\bibinfo {author} {\bibfnamefont {E.}~\bibnamefont
			{Witten}},\ }\href {https://doi.org/10.1016/0550-3213(82)90071-2} {\bibfield
		{journal} {\bibinfo  {journal} {Nucl. Phys. B}\ }\textbf {\bibinfo {volume}
			{202}},\ \bibinfo {pages} {253} (\bibinfo {year} {1982})}\BibitemShut
	{NoStop}%
	\bibitem [{\citenamefont {Correa}\ and\ \citenamefont
		{Plyushchay}(2007)}]{Correa2007p2493}%
	\BibitemOpen
	\bibfield  {author} {\bibinfo {author} {\bibfnamefont {F.}~\bibnamefont
			{Correa}}\ and\ \bibinfo {author} {\bibfnamefont {M.~S.}\ \bibnamefont
			{Plyushchay}},\ }\href {https://doi.org/10.1016/j.aop.2006.12.002} {\bibfield
		{journal} {\bibinfo  {journal} {Ann. Phys.}\ }\textbf {\bibinfo {volume}
			{322}},\ \bibinfo {pages} {2493} (\bibinfo {year} {2007})}\BibitemShut
	{NoStop}%
	\bibitem [{\citenamefont {Carey}\ \emph {et~al.}(2016)\citenamefont {Carey},
		\citenamefont {Gesztesy}, \citenamefont {Levitina},\ and\ \citenamefont
		{Sukochev}}]{Carey2016p881}%
	\BibitemOpen
	\bibfield  {author} {\bibinfo {author} {\bibfnamefont {A.}~\bibnamefont
			{Carey}}, \bibinfo {author} {\bibfnamefont {F.}~\bibnamefont {Gesztesy}},
		\bibinfo {author} {\bibfnamefont {G.}~\bibnamefont {Levitina}},\ and\
		\bibinfo {author} {\bibfnamefont {F.}~\bibnamefont {Sukochev}},\ }\href
	{https://doi.org/10.7153/oam-10-50} {\bibfield  {journal} {\bibinfo
			{journal} {Oper. Matrices}\ }\textbf {\bibinfo {volume} {10}},\ \bibinfo
		{pages} {881} (\bibinfo {year} {2016})}\BibitemShut {NoStop}%
	\bibitem [{\citenamefont {{NIST Digital Library of Mathematical
				Functions}}(2026)}]{NIST2026Mathieu}%
	\BibitemOpen
	\bibfield  {author} {\bibinfo {author} {\bibnamefont {{NIST Digital Library
					of Mathematical Functions}}},\ }\href@noop {} {\bibinfo {title} {{Mathieu}
			functions and {Hill}'s equation}},\ \bibinfo {howpublished}
	{\url{https://dlmf.nist.gov/28}} (\bibinfo {year} {2026}),\ \bibinfo {note}
	{release 1.2.8, 2026-09-15. Chapter 28, Secs.~28.2, 28.4, and
		28.8}\BibitemShut {NoStop}%
	\bibitem [{\citenamefont {Montgomery}(1967)}]{Montgomery1967p772}%
	\BibitemOpen
	\bibfield  {author} {\bibinfo {author} {\bibfnamefont {W.~D.}\ \bibnamefont
			{Montgomery}},\ }\href {https://doi.org/10.1364/JOSA.57.000772} {\bibfield
		{journal} {\bibinfo  {journal} {J. Opt. Soc. Am.}\ }\textbf {\bibinfo
			{volume} {57}},\ \bibinfo {pages} {772} (\bibinfo {year} {1967})}\BibitemShut
	{NoStop}%
	\bibitem [{\citenamefont {Hebri}\ \emph {et~al.}(2019)\citenamefont {Hebri},
		\citenamefont {Bagheri},\ and\ \citenamefont {Rasouli}}]{Hebri2019p4355}%
	\BibitemOpen
	\bibfield  {author} {\bibinfo {author} {\bibfnamefont {D.}~\bibnamefont
			{Hebri}}, \bibinfo {author} {\bibfnamefont {M.}~\bibnamefont {Bagheri}},\
		and\ \bibinfo {author} {\bibfnamefont {S.}~\bibnamefont {Rasouli}},\ }\href
	{https://doi.org/10.1364/OL.44.004355} {\bibfield  {journal} {\bibinfo
			{journal} {Opt. Lett.}\ }\textbf {\bibinfo {volume} {44}},\ \bibinfo {pages}
		{4355} (\bibinfo {year} {2019})}\BibitemShut {NoStop}%
\end{thebibliography}


\end{document}